\documentclass{aa}
\usepackage{txfonts}
\usepackage{caption}
\usepackage{amssymb, amsmath}
\usepackage{array}
\usepackage{color}
\usepackage{graphicx}
\usepackage{hyperref}
 \newcommand{\mathbfss}[1]{\textbf{\textsf{#1}}}

\begin{document}

\title{Generalised model-independent characterisation of strong gravitational lenses VII: impact of source properties and higher-order lens properties on the local lens reconstruction}
\titlerunning{Impact of source properties and flexion}
\author{Jenny Wagner\inst{1}} 
\institute{\url{thegravitygrinch.blogspot.com}, 
\email{thegravitygrinch@gmail.com}
}
\date{Received XX; accepted XX}

\abstract{
We investigate the impact of higher-order gravitational lens properties and properties of the background source on our approach to directly infer local lens properties from observables in multiple images of strong gravitationally lensed extended, static background sources developed in papers I to VI. 
As the degeneracy between local lens and source properties only allows to determine \emph{relative} local lens properties between the multiple image positions, we cannot distinguish common scalings and distortions caused by lensing from intrinsic source characteristics. 
The consequences of this degeneracy for lens modelling and our approach and ways to break it are detailed here.
We also set up quantitative measures around the critical curve to find clear limits on the validity of the approximation that source properties are negligible to infer local lens properties at critical points. 
The impact of the source on the local lens properties depends on the reduced shear at the image position and the amplitude and orientation of the source ellipticity, as we derive in this paper.
Similarly, we investigate the role of third-order lens properties (flexion), in two galaxy-cluster simulations and in the \texttt{Lenstool}-reconstruction of the galaxy-cluster lens CL0024. 
In all three cases, we find that flexion is negligible in over 90\% of all pixels of the lensing region for our current imprecision of local lens properties of about 10\%. 
Decreasing the imprecision to 2\%, higher-order terms start to play a role, especially in regions with shear components close to zero. 
}
\maketitle

\section{Introduction}
\label{sec:introduction}

\subsection{Brief outline of parts I to VI}
\label{sec:previous_work}

The paper series of \cite{bib:Wagner1}, \cite{bib:Wagner2}, \cite{bib:Wagner3}, \cite{bib:Wagner4}, \cite{bib:Wagner5}, and \cite{bib:Wagner6} develops a new method to infer local properties of strong gravitational lenses in the vicinity of critical curves without employing a mass density profile for the gravitational lens.
As detailed in \cite{bib:Wagner5}, we also freed the approach from the assumption of a specific parametrisation of a cosmological model for a homogeneous, isotropic universe in which a metric theory describes gravitational interactions.
\cite{bib:Wagner_universe} summarises the current state of development and gives a synopsis of the observational examples to which our approach has been applied.

Details about the applications can be found in \cite{bib:Wagner_cluster}, \cite{bib:Wagner_frb}, \cite{bib:Wagner_quasar}, which also show that the resulting local lens properties of our approach are in agreement with state-of-the-art lens reconstruction algorithms that employ superpositions of parametric mass density profiles, like \texttt{Lenstool} (\cite{bib:Kneib}, \cite{bib:Jullo}) and \texttt{Lensmodel} (\cite{bib:Keeton1}, \cite{bib:Keeton2}) or those algorithms that use series of weighted basis functions, like \texttt{Grale} (\cite{bib:Liesenborgs1}, \cite{bib:Liesenborgs2}) and \texttt{PixeLens} (\cite{bib:Saha}) to describe the distribution of the total gravitationally lensing mass within a whole region of interest.
\cite{bib:Griffiths} shows an example case that can only be tackled by our approach because merely three multiple images of a single background galaxy have been identified in the lensing galaxy cluster. 

We also showed that the comparison between the complementary approaches allows us to corroborate or reject additional model assumptions, like the light-traces-mass assumption and to evaluate the influence of lens-model regularisations in the vicinity of the multiple images.
Since our approach uses observables from the multiple images to reconstruct local lens properties, it is highly useful for the reconstruction of local, small-scale details in the mass density profile.
These small-scale details require computationally intensive calculations when global lens reconstructions are employed. 

With increasing number of multiple images observed by future telescopes, see e.g.~\cite{bib:Ghosh}, our observation-based approach can equip an increasing lensing region with lens-model-independent properties. 
Thereby, we will be able to replace assumption-based lens properties by evidence-based properties in a growing lensing region.

In this part of the series, we investigate the impact that intrinsic source properties of extended, multiply-imaged background objects with static brightness profiles, like galaxies, have on the reconstruction of the local lens properties. 
As intrinsic source properties are degenerate with local lens properties (see paper IV for details, \cite{bib:Wagner4}), we also investigate the influence of higher-order lensing effects. 
Identifying the amount and distribution of lensing regions in which third-order flexion exceeds a given threshold relative to second-order convergence and shear, we constrain its impact on our local lens reconstruction and on global lens reconstructions at the same time.
As in all previous papers of this series and mostly assumed in the lens reconstruction algorithms mentioned above, we assume an effective lens description in which a single lens plane with a two-dimensional mass density causes all light deflections.

\subsection{Prerequisites and related work}
\label{sec:related_work}

To describe light propagation and its deflections, the gravitational lensing formalism, as detailed e.g.~in \cite{bib:SEF}, used to be based on infinitesimal light beams and infinitesimal sources established in \cite{bib:Sachs}.
``Infinitesimal'' means that the extension of the beam orthogonal to its propagation direction fulfils three approximations.
The first one being that its cross section is small compared to the curvature of the propagating wave front, meaning that its extension on the sky is small compared to its distance from us as observers. 
The second one being that the deflecting gravitational field is caused by masses moving at speeds much less than the speed of light and that the field is weak, such that spacetime is flat over the cross section of the light beam and the deflecting gravitational potential is Newtonian.
The third one being that the spacetime curvature remains constant over the cross section of the beam, which guarantees the geodesic deviation equation to hold across the beam.

Strong and weak gravitational lensing fulfil the first two criteria to a good approximation.
Generalising the standard gravitational lensing formalism to extended, non-infinitesimal light beams requires an investigation of the third criterion (\cite{bib:Fleury1}).
Assuming that the light beam encloses several discrete mass clumps, or a continuous inhomogeneous excess mass density, the third assumption is not valid anymore. 
Observable effects include micro-lensing on top of the homogeneous background density of the universe, see e.g.~\cite{bib:Kains} or \cite{bib:Mroz}, micro-lensing on top of a larger-scale gravitational lens like a galaxy or a galaxy cluster, see e.g.~\cite{bib:Millon}, \cite{bib:Chen}, or lensing effects caused by small-scale perturbing mass densities on top of larger-scale gravitational lenses like galaxies, see e.g.~\cite{bib:Vegetti}, \cite{bib:Walls}.

The influence of intrinsic properties of extended background sources on the reconstruction of the gravitational lens can be degenerate with the influence of small-scale light deflecting structures along the path of the extended light bundle. 
This degeneracy is caused by the fact that, for galaxy-scale and galaxy-cluster-scale gravitational lenses, the extended background source is unobservable, as well as the dark matter part of the deflecting mass density distribution cannot be directly determined.
We show how this degeneracy enters the lensing formalism and investigate ways to break it in Section~\ref{sec:degeneracy1}.

Subsequently, we employ extended light bundles obtained by means of a Taylor expansion around an infinitesimal light bundle and we will assume that the curvature is slowly and smoothly varying, such that the extended light bundles experience only small deviations from a constant curvature.
This approximation has already been successfully employed in many works, see e.g.~\cite{bib:SEF}, \cite{bib:Petters}, \cite{bib:Kochanek}, in the previous parts of this paper series, and for the applications mentioned in Section~\ref{sec:previous_work}. 
So far, our observational investigations detailed in \cite{bib:Wagner_cluster}, \cite{bib:Wagner_quasar}, and \cite{bib:Griffiths} showed that the current measurement precision and observational resolution corroborate the validity of this assumption. 
We found that the convergence and the reduced shear are constant to a good approximation over the area of multiple image brightness profiles, so that we can resort to the influence of source properties on these quantities in Section~\ref{sec:source_props}.

Employing the simulated mass density profiles ARES and HERA galaxy clusters from $N$-body simulations from \cite{bib:Meneghetti}, we investigate whether our approach to only focus on convergence and shear is supported by state-of-the-art realistic gravitational lens simulations in Section~\ref{sec:application}. 
The simulated data also allow us to determine the regions around the multiple images and around the critical curve where intrinsic source ellipticities are negligible -- an assumption used in \cite{bib:Wagner1} when approximating the location and slope of the critical curve in the vicinity of multiple images.
Furthermore, the simulations allow us to estimate how often and where we expect the degeneracies detailed in Section~\ref{sec:degeneracy1} to occur in observations. 
To compare the simulated results with observation-based ones, we analogously evaluate the global lens reconstruction of CL0024+1654, CL0024 for short, that we obtained in \cite{bib:Wagner_cluster} using the reconstruction algorithm \texttt{Lenstool}. 
%
Section~\ref{sec:conclusion} summarises all outcomes of this part. 

\subsection{Notation}
\label{sec:notation}

To make this part self-consistent, we briefly revise the notation and variables of the standard single-plane gravitational lensing formalism we use in the entire paper series.
Based on the notation in \cite{bib:SEF}, we define the following quantities in Cartesian coordinates:
\begin{itemize}
\item $\boldsymbol{y} = (y_1, y_2)$ as the angular position of a background source located at redshift $z_\mathrm{s}$ along the line of sight,
\item $\boldsymbol{x}_J = (x_{J1}, x_{J2})$ as the angular position of the multiple image $J$ at redshift $z_\mathrm{l}$ along the line of sight,
\item $\psi(\boldsymbol{x})$ as the deflection potential, i.e.~the Newtonian gravitational potential of all masses between the source and the observer, projected into the lens plane located at $z_\mathrm{l}$ along the line of sight and scaled to be an angular (dimensionless) quantity like $\boldsymbol{y}$ and $\boldsymbol{x}$.
\end{itemize}

\section{Intrinsic source properties and small-scale lens perturbations}
\label{sec:degeneracy1}

\subsection{The source position - deflection potential degeneracy}
\label{sec:spdpd}

In \cite{bib:Wagner4}, we derived that the time delay difference measured between two multiple images of a time-varying source remains invariant if we transform the deflection potential of the lens and the position of the source as
\begin{equation}
\delta \psi (\boldsymbol{x}_I) - \delta \psi (\boldsymbol{x}_J) = - \left( \boldsymbol{x}_I - \boldsymbol{x}_J \right)^\top \delta \boldsymbol{y} \;.
\label{eq:deg}
\end{equation}
$\delta \psi(\boldsymbol{x})$ denotes the change in the deflection potential from one deflection potential $\psi(\boldsymbol{x})$ to another $\tilde{\psi}(\boldsymbol{x})$ at the angular position $\boldsymbol{x}$ in the lens plane. Analogously, $\delta \boldsymbol{y}$ denotes the change in the angular source position from $\boldsymbol{y}$ to $\tilde{\boldsymbol{y}}$ in the source plane.
Even when fixing the time delay difference between two multiple images, the degeneracy in Equation~\eqref{eq:deg} remains for arbitrary functions to change the local deflection potential, $\delta \psi(\boldsymbol{x})$, and arbitrary relocation functions $\delta \boldsymbol{y}$.
The arbitrariness in the choice of the transformations of the deflection potential and the source position holds, as long as $\psi(\boldsymbol{x})$ and $\boldsymbol{y}$ are considered as independent variables. 
Imposing the lens equation, makes these variables dependent:
\begin{equation}
\delta \boldsymbol{y} = \nabla_{\boldsymbol{x}} \delta \psi(\boldsymbol{x}) \;, \quad \text{for all multiple image positions} \; \boldsymbol{x} \;.
\end{equation}
Consequently, $\delta \boldsymbol{y}$ can only be a translation in the source position, so that the transformation of $\psi(\boldsymbol{x})$ is a linear function at the multiple image positions. 
As already emphasised in \cite{bib:Wagner4}, all statements refer to the multiple image positions within the extension of their observable brightness profiles and are not meant to be global transformations of the lens or source plane. 

Physically, Equation~\eqref{eq:deg} can be interpreted that local small-scale changes in the deflection potential and relocations of the common source are degenerate with each other, as long as the source position remains unobservable and there is no direct measure to determine the deflection potential at the multiple image positions, as also derived in \cite{bib:Fleury2}.
Consequently, the absolute position of the source within the source plane cannot be determined.
Measuring the time delay difference cannot break this degeneracy alone because it only constrains the difference in the Fermat potential between the two multiple images, see also \cite{bib:Liesenborgs3}. 
Additional model assumptions, e.g. about the global lens structure or the smoothness of the potential, can contribute to break the degeneracy of Equation~\eqref{eq:deg} (\cite{bib:Wagner6}).
Not fixing the time delay difference by observations and allowing it to vary as well, the space of degeneracies increases even more, as detailed in \cite{bib:Wagner4}. 

When employing model-based lens reconstructions, we can interpret Equation~\eqref{eq:deg} differently.
If we assume that each image $I$ has its own back-projection $\boldsymbol{y}_I$ and $\delta \boldsymbol{y}_I$ denotes a relocation of the back-projection $\boldsymbol{y}_I$ from a common source position $\boldsymbol{y}$ of all back-projections, then Equation~\eqref{eq:deg} reads
\begin{equation}
\delta \psi (\boldsymbol{x}_I) - \delta \psi (\boldsymbol{x}_J) = - \left( \boldsymbol{x}_I - \boldsymbol{x}_J \right)^\top \tfrac12 \left( \delta \boldsymbol{y}_I + \delta \boldsymbol{y}_J \right) \;.
\label{eq:deg2}
\end{equation} 
By ``fine-tuning'' the global lens model with small-scale adjustments at each image position $I$, it is always possible to find locally confined $\delta \psi (\boldsymbol{x})$ such that $\delta \boldsymbol{y}_I + \delta \boldsymbol{y}_J = 0$.
In this case, the right adjustments in the lens reconstruction can force all back-projections to overlap in one common background source.
Thus, the \emph{relative} location of the common background source can be optimised, as is done in the lens reconstruction algorithms, see e.g.~the reconstruction algorithms mentioned in Section~\ref{sec:introduction}. 
Yet, as already stated above, the \emph{absolute} position of this common background source is degenerate with the local small-scale corrections to the deflection potential.

As Equation~\eqref{eq:deg2} only weakly constrains the invariance transformations $\delta \psi(\boldsymbol{x})$ and $\delta \boldsymbol{y}$, model-based reconstructed source morphologies can be altered beyond an overall translation of the common source reconstruction in the source plane.
Preventing model-based source and lens reconstructions to be subject to this degeneracy, model-based methods need to set up a constraint on the lowest allowed fine-tuning of the deflection potential or give a lower limit at which overlap of back-projected multiple images the common source position can be constrained by the data.

To break the degeneracies in Equations~\eqref{eq:deg} and \eqref{eq:deg2}, we can add assumptions about the small-scale lens properties in terms of lens models or assumptions about the source properties and its position, which has been the standard way to determine source or small-scale lens properties so far. 
In contrast, we investigate which lens and source properties can be inferred without employing lens model assumptions.

Since Equations~\eqref{eq:deg} and \eqref{eq:deg2} also contain the image positions, \cite{bib:Birrer} calculate the constraining power of back-projections of imprecisely measured image positions onto the source plane assuming a known lens model.
Using example image configurations of a singular isothermal ellipse with external shear, they show that, even with perfect knowledge of the lens, the common source position can only be constrained to a small region due to the finite measurement precision of the image positions. 
This measurement uncertainty needs to be taken into account in the source reconstruction in addition to the invariance transformations $\delta \psi(\boldsymbol{x})$ and $\delta \boldsymbol{y}$, which represent the remaining freedom in the lens and source reconstruction not constrained by the observables. 

\subsection{Infinitesimal-beam approximation}
\label{sec:infinitesimal_beam}

At first, we need to investigate at which point the prerequisite of infinitesimal beams breaks down and it is not sufficient anymore to characterise the impact of the gravitational lens by means of the distortion matrix, i.e.~by the enlarging or shrinking convergence and the shear distortions. 

Hence, we expand the lens mapping 
\begin{equation}
\boldsymbol{y}_0 = \boldsymbol{x}_0 - \nabla_{\boldsymbol{x}} \psi(\boldsymbol{x}_0)  = \boldsymbol{x}_0 - \boldsymbol{\alpha}(\boldsymbol{x}_0)
\label{eq:le}
\end{equation}
around a corresponding source and image position pair $(\boldsymbol{y}_0, \boldsymbol{x}_0)$ to higher order. 
Without loss of generality, we assume that $\boldsymbol{y}_0$ and $\boldsymbol{x}_0$ are the origins of the coordinate systems in the source and lens planes, respectively.
A convenient choice of $\boldsymbol{y}_0$ is the position within the brightness profile of the source object that shows maximum intensity.
If $\boldsymbol{y}_0$ is the source position with maximum intensity, $I(\boldsymbol{y}_0)$ with $\nabla_{\boldsymbol{y}} I(\boldsymbol{y}_0)=0$, $\boldsymbol{x}_0$ will also be an intensity maximum in the lens plane due to the conservation of surface brightness
\begin{equation}
\nabla_{\boldsymbol{x}} I(\boldsymbol{x}_0) = \mathbfss{A}(\boldsymbol{x}_0,\boldsymbol{y}_0) \nabla_{\boldsymbol{y}} I(\boldsymbol{y}_0(\boldsymbol{x}_0)) = 0 \;,
\end{equation}
$\mathbfss{A} \equiv \mathbfss{A}(\boldsymbol{x},\boldsymbol{y})$ denoting the distortion matrix with entries
\begin{equation}
A_{ij} = \dfrac{\partial y_i}{\partial x_j} = \dfrac{\partial}{\partial x_j} \left( x_i - \dfrac{\partial}{\partial x_i} \psi(\boldsymbol{x}) \right) \;, \quad i,j=1,2 \;,
\label{eq:A}
\end{equation}
evaluated at $(\boldsymbol{y}_0, \boldsymbol{x}_0)$.
Deriving $\mathbfss{A}$ with respect to $\boldsymbol{x}$, we arrive at the flexion terms $\mathbfss{D} \equiv \mathbfss{D}(\boldsymbol{x},\boldsymbol{y})$ with the entries
\begin{equation}
D_{ijk} =   \dfrac{\partial^2 y_i}{\partial x_j \partial x_k} \;, \quad i,j, k=1,2\;. 
\label{eq:D}
\end{equation}
\cite{bib:Bacon} give an encompassing overview of the definitions, interpretations, and effects of the terms in $\mathbfss{A}$ and $\mathbfss{D}$.
We only employ the physical interpretations in terms of convergence $\kappa(\boldsymbol{x})$ and shear $\boldsymbol{\gamma}(\boldsymbol{x})=(\gamma_1(\boldsymbol{x}),\gamma_2(\boldsymbol{x}))$
\begin{equation}
\mathbfss{A} = \left( \begin{matrix} 1- \kappa(\boldsymbol{x}) & 0 \\ 0 & 1-\kappa(\boldsymbol{x}) \end{matrix} \right) - \left( \begin{matrix} \gamma_1(\boldsymbol{x}) & \gamma_2(\boldsymbol{x}) \\ \gamma_2(\boldsymbol{x}) & -\gamma_1(\boldsymbol{x}) \end{matrix} \right) \;,
\label{eq:A2}
\end{equation}
such that the distortion matrix is interpreted as a scaling by $(1-\kappa(\boldsymbol{x}))$ (first term) and a distortion by the symmetric shear matrix (second term). Consequently, $\mathbfss{D}$, containing their derivatives, represents the changes in convergence and shear. 

Using Equations~\eqref{eq:A} and \eqref{eq:D}, the Taylor expansion of the lens mapping to second order around $(\boldsymbol{y}_0, \boldsymbol{x}_0)=(\boldsymbol{0},\boldsymbol{0})$ reads
\begin{equation}
y_i = \sum \limits_{j=1}^{2} A_{ij} x_j + \dfrac12 \sum \limits_{j=1}^2 \sum \limits_{k=1}^2 D_{ijk} x_j x_k \;, \quad i=1,2\;.
\label{eq:Taylor}
\end{equation} 
The approximation that the distortion matrix is sufficient to characterise the lensing properties in the vicinity of $\boldsymbol{x}_0$ breaks down if $\boldsymbol{x}$ is so far from $\boldsymbol{x}_0$ that $\mathbfss{A}(\boldsymbol{x},\boldsymbol{y}) \approx \mathbfss{A}(\boldsymbol{x}_0,\boldsymbol{y}_0)$ is not valid anymore and the second term on the right-hand side in Equation~\eqref{eq:Taylor} needs to be included. 
Expressing $\mathbfss{D}$ and $\mathbfss{A}$ in terms of convergence and shear, we can thus neglect the derivatives of $\kappa(\boldsymbol{x})$ and $\boldsymbol{\gamma}(\boldsymbol{x})$ in Equation~\eqref{eq:Taylor} if
\begin{equation}
\dfrac12 \left| \sum \limits_{k=1}^2 \dfrac{\partial \kappa(\boldsymbol{x})}{\partial x_k} x_k \right| \ll \kappa(\boldsymbol{x}) \, \wedge \, \dfrac12 \left| \sum \limits_{k=1}^2 \dfrac{\partial \gamma_{i}(\boldsymbol{x})}{\partial x_k} x_k \right| \ll |\gamma_{i}(\boldsymbol{x})|
\end{equation}
with $i,j=1,2$.
To determine whether these relations hold in practice, we require that
\begin{equation}
\dfrac12 \left| \sum \limits_{k=1}^2 \dfrac{\partial \kappa(\boldsymbol{x})}{\partial x_k} x_k \right| \le \delta_\kappa \kappa(\boldsymbol{x}) \, \wedge \, \dfrac12 \left| \sum \limits_{k=1}^2 \dfrac{\partial \gamma_{i}(\boldsymbol{x})}{\partial x_k} x_k \right| \le \delta_{\gamma_i} |\gamma_{i}(\boldsymbol{x})|
\label{eq:neglect_D}
\end{equation} 
for some pre-defined threshold parameters $0 < \delta_\kappa, \delta_{\gamma_i} < 1$.

Simulated gravitational lenses or lens reconstructions from observational data usually give $\kappa$- and $\boldsymbol{\gamma}$-maps over the entire lensing region.
Calculating maps of the derivatives of $\kappa$ and $\boldsymbol{\gamma}$, Equation~\eqref{eq:neglect_D} allows us to determine the lensing regions in which $\mathbfss{A}$, i.e.~$\kappa(\boldsymbol{x})$ and $\boldsymbol{\gamma}(\boldsymbol{x})$, can be approximated as being constant and used to fully characterise the local lens properties in these lensing regions.
Section~\ref{sec:application} shows two example cases how these regions can be determined in realistically simulated galaxy-cluster lenses and in a lens-model-based reconstruction of CL0024.

If a low measurement precision or simulation resolution does not allow us to constrain the entries of $\mathbfss{D}$, the approximation with a constant distortion matrix for the entire area covered by a multiple image is sufficient. 
The expected use-case for this approximation are multiple images whose extension is small compared to the (effective) Einstein radius of the lens, like, for instance, multiple images generated by galaxy-cluster lenses.
As detailed in \cite{bib:Wagner3}, giant arcs and Einstein ring images around galaxy-scale lenses do not fall into this category.

\subsection{Lensing regions with negligible distortions or scalings}
\label{sec:no_A}

In \cite{bib:Wagner2} and \cite{bib:Wagner_cluster}, we systematically investigated the reconstruction quality of the ratios of convergences and reduced shear for all positions of $n$ multiple images from a common background source
\begin{equation}
f_{IJ} \equiv \dfrac{1-\kappa(\boldsymbol{x_J})}{1-\kappa(\boldsymbol{x_I})} \;, \quad \boldsymbol{g}_I \equiv \dfrac{\boldsymbol{\gamma}(\boldsymbol{x_I})}{1-\kappa(\boldsymbol{x_I})} \;, \quad I,J=1,2,...,n
\label{eq:local_lens_props}
\end{equation}
by our approach. 
We found that increasing the area covered by the distorted brightness profile of a multiple image increases the reconstruction precision of its local lens properties.
Concerning the shape of these brightness profiles, highly elongated shapes result in a worse precision for the local lens properties than less elliptical images. 
Hence, given a set of observed multiple images, we can select the one with the most precise local lens properties to reconstruct the source morphology.
Further details about the impact of the degeneracy in Equation~\eqref{eq:deg} on the source and local lens reconstruction can be found in Section~\ref{sec:source_props}. 

Complementary to this ansatz, we can also search for those lensing regions fulfilling the infinitesimal-beam approximation (Equation~\eqref{eq:neglect_D}) and that show maximum convergence at minimum shear, as already noted in \cite{bib:Williams} dubbing this kind of images HMU images, short form of highly-magnified undistorted images. 
Multiple images located in these regions are optimal observables to reconstruct the source morphology.
The minimum shear guarantees that the degeneracy between local lens and intrinsic source properties is broken up to the overall scaling factor of the enlarging convergence. 
The latter allows us to inspect details in the source that would be unobservable without the gravitational lensing effect. 
\cite{bib:Zitrin} discovered an example case in MACS J1149.5+2223, which happens to be the spiral galaxy hosting Supernova Refsdal \cite{bib:Kelly}. 
Section~\ref{sec:application} shows these regions for two simulated example galaxy-cluster lenses.
An analogous analysis can be performed for global lens-model-based lens reconstructions.

Beyond source morphology reconstructions, \cite{bib:Morioka} investigated the statistics of such HMUs, called GRAMORs (GRavitationally highly magnified yet MORphologically regular images) in their work, to constrain cosmological parameters. 
They find that HMUs/GRAMORs put tighter constraints on the latter than giant arcs. 

More difficult to find are images in regions with $\kappa\approx1$ which are not scaled but only distorted. 
Yet, these images are invariant with respect to any mass-sheet transform, and therefore could help to break the mass-sheet degeneracy in lens reconstructions, if they can be reliably identified.
Thus multiple images in these special regions, without shear or with $\kappa \approx 1$, can contribute to break the degeneracies between lens and source properties.  

\section{Local lens properties from multiply-imaged sources with non-negligible intrinsic properties}
\label{sec:source_props}

In the following, we assume the infinitesimal-beam approximation to hold and only take into account convergence and shear as local lens properties. 
Calculating the local lens properties at the positions of multiple images from a linear transformation of these images onto each other was first mentioned in \cite{bib:Gorenstein1} and extended to probe a rotation degree of freedom in \cite{bib:Pen}. 
We worked the case of \cite{bib:Gorenstein1} out in detail in \cite{bib:Wagner2} and \cite{bib:Wagner_cluster} and tested it in a simulation and a galaxy-cluster-scale example.
Beyond that, extending the employed Taylor expansion from the areas in the lens plane covered by the multiple images to the critical curves, local approximations to the critical curves can be determined as detailed in \cite{bib:Wagner1}. 
All of these approaches assumed that the source properties are negligible, in particular, that all observable ellipticities of multiple images are caused by the lens.
Therefore, the approach is extended in this section taking intrinsic source properties into account.  

\subsection{Local lens properties from transformations between multiple images}
\label{sec:part1}

We transform a multiple image $I$ onto an image $J$ from the same background source by means of an affine transformation $\mathbfss{T}_{IJ}$. 
In \cite{bib:Wagner2}, we set up it up as
\begin{equation}
\mathbfss{T}_{IJ} = \mathbfss{A}_J^{-1} \,\mathbfss{A}_I \;,
\label{eq:T}
\end{equation}
i.e.~the product of the respective distortion matrices of image $I$ and $J$.
$\mathbfss{A}_I$ and $\mathbfss{A}_J$ are assumed to be constant over the extensions of the brightness profiles of the multiple images, so that their entries can be abbreviated as $\kappa_I$, $\kappa_J$ and $\boldsymbol{\gamma}_I$, $\boldsymbol{\gamma}_J$. 
Since the invertible affine transformations form a group, $\mathbfss{T}_{IJ}$ as a product of affine transformations is an affine transformation as well.
As such, we can decompose it again into its scaling and shear parts. 
Appendix~\ref{app:T} details the calculation to arrive at 
\begin{align}
\mathbfss{T}_{IJ} =& \left(1-\dfrac{\kappa_I - \kappa_J}{1-\kappa_J}  \right)  \left(\begin{matrix}1 & 0 \\ 0& 1 \end{matrix} \right) \nonumber \\
 &- (1-\kappa_I)\mathbfss{A}_J^{-1} \left( \begin{matrix} g_{I,1}-g_{J,1} & g_{I,2}-g_{J,2} \\ g_{I,2}-g_{J,2} & -(g_{I,1}-g_{J,1}) \end{matrix} \right) \;.
 \label{eq:T_map}
\end{align}
The mapping of two images onto each other thus consists of an identity transformation, followed by a scaling with a convergence term proportional to the difference between the convergences, followed by a scaled shear transformation proportional to the shear difference between the two images.
If the local lens properties are the same for both images, $\mathbfss{T}_{IJ}$ reduces to the identity transformation as expected from Equation~\eqref{eq:T} for $\mathbfss{A}_I=\mathbfss{A}_J$.
Hence, linear transformations between image pairs can only determine the \emph{differences} of the local lens properties between the two image locations $I$ and $J$. 

Since the transformation between the images only incorporates differences between the local lens properties, the identity transformation in the first term contains an additional degeneracy between local lens properties and their intrinsic source properties. Assuming a transformation in the source plane denoted by $\mathbfss{E}(\boldsymbol{y})$, any additional local lens distortion $\mathbfss{B}$ (as defined by Equation~\eqref{eq:A2}) on top of the distortion already assumed in $\mathbfss{A}$ compensating the transformation $\mathbfss{E}$ of the source, such that
\begin{equation}
\mathbfss{B}_J^{-1}\mathbfss{E}(\boldsymbol{y}) \mathbfss{B}_I = \left( \begin{matrix} 1 & 0 \\ 0 & 1 \end{matrix} \right) \;,
\label{eq:deg3}
\end{equation}
remains undetected while inferring the local lens properties of Equation~\eqref{eq:local_lens_props} by means of Equation~\eqref{eq:T_map}.
The same result can also be inserted into the product of distortion matrices.
Inserting an identity transformation and relating it with Equation~\eqref{eq:deg3}, we arrive at
\begin{equation}
\mathbfss{T}_{IJ} = \mathbfss{A}_J^{-1} \left( \begin{matrix} 1 & 0 \\ 0 & 1 \end{matrix} \right) \mathbfss{A}_I = \mathbfss{A}_J^{-1} \mathbfss{B}_J^{-1}\mathbfss{E}(\boldsymbol{y}) \mathbfss{B}_I \mathbfss{A}_I \;,
\label{eq:deg_s}
\end{equation}
which is the linearised version of Equation~\eqref{eq:deg} stating that local lens properties are degenerate with a transformation of the source, as also discussed in \cite{bib:Wagner4} to distinguish Equations~\eqref{eq:deg} and \eqref{eq:deg_s} from the global source position transformation of \cite{bib:Schneider_2014}.
This degeneracy can only be broken if we know the absolute brightness and the brightness profile of the background source, as detailed in \cite{bib:Wagner_universe}. 

For all source objects that cannot be resolved by the magnification of the gravitational lens, i.e. in the limit of infinitesimal beams whose cross sections are constant brightness profiles, the degeneracy can be broken up to an overall scaling factor.
This approximation certainly holds for pixel-wise source reconstruction techniques as, for instance the approaches developed in \cite{bib:Suyu_source}, \cite{bib:Vegetti_source}, and the pixel-wise source reconstructions of the lens modelling algorithms mentioned in Section~\ref{sec:introduction}. 

Another common way to break the degeneracy in Equation~\eqref{eq:deg3} is to identify matching features in the brightness profiles of all multiple images and assume them to be intrinsic source properties.
Yet, Equation~\eqref{eq:T_map} only determines differences in local lensing properties, so that the observed matching features may directly be attributed to the source or may be a combination of intrinsic source properties and a common distortion caused by the gravitational lens. 
The assumption that matching brightness features are intrinsic source properties is successfully applied to the multiple images in the galaxy-cluster lens CL0024 (see \cite{bib:Wagner_cluster} and references therein) and to Hamilton's Object (see \cite{bib:Griffiths}), leading to self-consistent results in both cases.
For the multiple images generated by the galaxy lens B0128 the matching of the salient features across all multiple images is problematic. 
As detailed in \cite{bib:Wagner_quasar} and references therein, the matching of salient brightness features in accordance with leading order lens theory returns inconsistent local lens properties. 
Recently, \cite{bib:Ivison} further investigated the reality of such salient features detected in interferometric sub-millimeter observations, using new, high-resolution ALMA observations. 
Discovering that the brightness blobs previously detected in SMM J21352-0102 (the ``Cosmic Eyelash") are spurious, it may be possible that other analogous observations, like the brightness blobs in the multiple images in B0128, are also observational artefacts.
Hence, matching the brightness profiles of multiple images with each other may result in incorrect matchings and entail biased lens and source reconstructions due to the degeneracy in the formalism and due to uncertainties and misinterpretations of observables.

We can thus conclude that the degeneracy between the source and local lens properties is only broken for standardisable extended background objects with known absolute brightness and brightness profile\footnote{Assuming that we cannot directly observe the source, as would be possible in transient lensing configurations like in plasma or micro lensing.}.
If the background object is unknown, the least degenerate local lens properties are obtained when matching infinitesimal beams of constant cross-section onto each other. 

\subsection{Local lens properties close to the critical curve}
\label{sec:part2}

Based on \cite{bib:SEF}, \cite{bib:Wagner0} and \cite{bib:Wagner1} developed a method to infer local properties of the critical curve from observables in multiple images alone. 
Using a Taylor expansion of the deflection potential around a critical point, the critical point itself and the slope of the critical curve in its vicinity were the local lens properties constrained by static, extended multiple images. 
The derivations of these equations assumed that the entire distortion and scaling of the multiple images were caused by the convergence and shear in the distortion matrix. 

To investigate the influence of the source properties, we convert the local lens properties in terms of derivatives of the Fermat potential into (reduced) convergence- and shear-related quantities, as they are easier to interpret physically.  
Rewriting Equation~\eqref{eq:A2}, we split the distortion matrix into an overall scaling proportional to $\kappa \equiv \kappa(\boldsymbol{x})$, an amplitude $|\boldsymbol{g}| \equiv |\boldsymbol{g}(\boldsymbol{x})|$, and a direction $\varphi \equiv \varphi(\boldsymbol{x})$ of reduced shear
\begin{equation}
\mathbfss{A} =  \left( 1- \kappa \right) \left( \begin{matrix} 1 - |\boldsymbol{g}| \cos\left( 2 \varphi \right)& \phantom{1}- |\boldsymbol{g}| \sin\left( 2 \varphi \right) \\ \phantom{1}- |\boldsymbol{g}| \sin\left( 2 \varphi \right) & 1 + |\boldsymbol{g}| \cos\left( 2 \varphi \right)\end{matrix} \right) \;,
\label{eq:A3}
\end{equation}
in which $2 \varphi$ represents the fact that the shear is a spin-2 quantity. 
Furthermore, let $a_\mathrm{s}$ and $b_\mathrm{s}$ denote the lengths of the semi-major and the semi-minor axes of the source quadrupole and let the latter be arbitrarily oriented in the source plane by an angle $\vartheta_\mathrm{s}$ with respect to the $y_1$-axis, analogously to the distortion matrix.

\begin{figure*}[ht]
\centering
\includegraphics[width=0.45\textwidth]{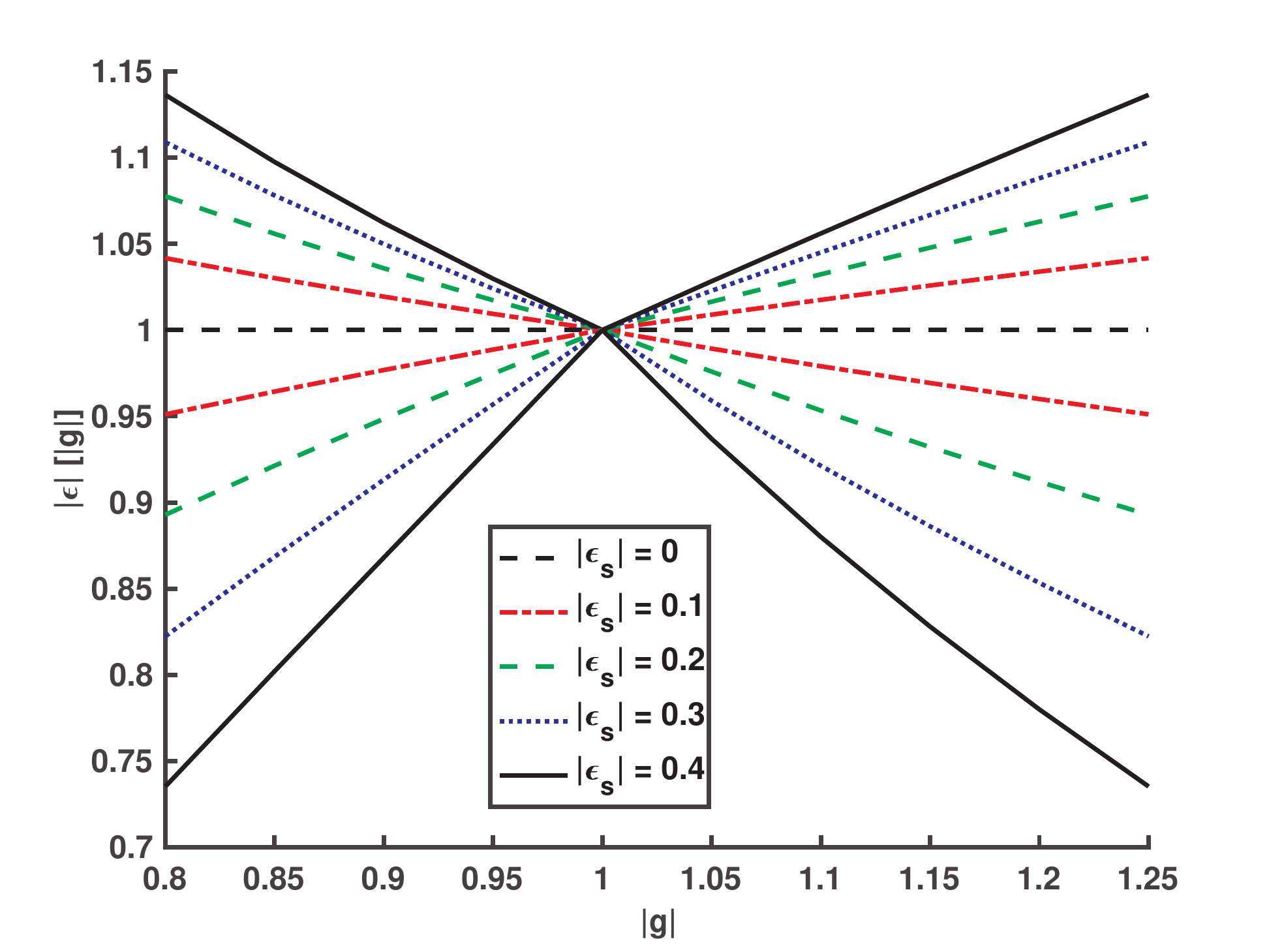} \hspace{4ex}
\includegraphics[width=0.45\textwidth]{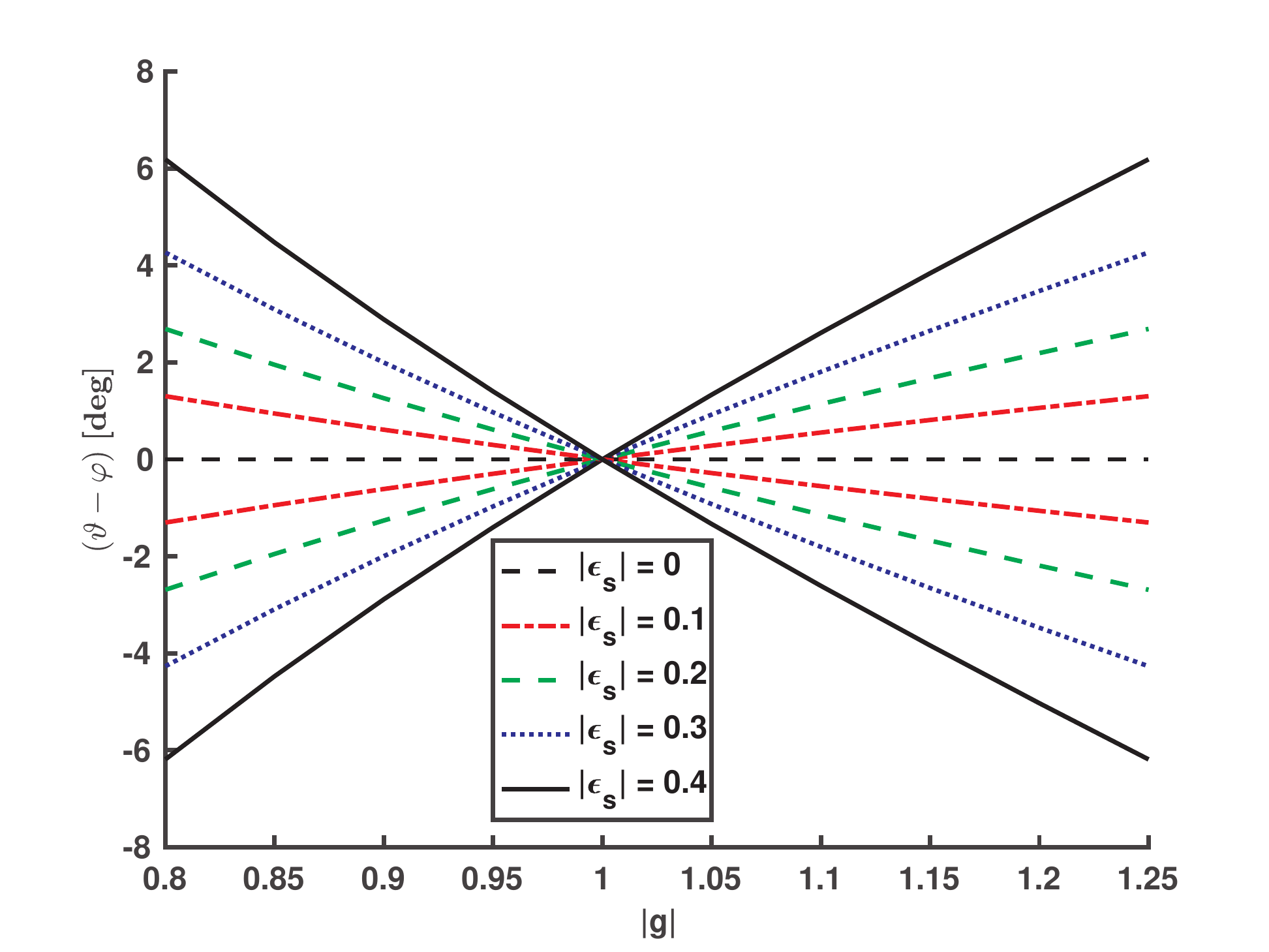}
\caption{Maximum influence of source properties in terms of the complex source quadrupole ellipticity $\boldsymbol{\epsilon}_\mathrm{s}$ on the inference of the complex (reduced) shear $\boldsymbol{g}$ when observing the complex image quadrupole ellipticity $\boldsymbol{\epsilon}$ close to the critical curve at which $|\boldsymbol{g}|=1$. 
The left plot shows the absolute value of Equations~\eqref{eq:eps1} and \eqref{eq:eps2} for source quadrupoles aligned parallel and anti-parallel to the (reduced) shear. 
$|\boldsymbol{\epsilon}|$ is larger than $|\boldsymbol{g}|$ when $\varphi - \vartheta_\mathrm{s} = 0$, while the opposite occurs when $\varphi - \vartheta_\mathrm{s} = \pi/2$. 
The right plot shows the relative orientation between the measured orientation $\vartheta$ of the image quadrupole and the orientation of the (reduced) shear $\varphi$ for these source ellipticities. Here, as plotted, the maximum deviation occurs for $\varphi - \vartheta_\mathrm{s} = \pm \pi/4$.}
\label{fig:alignment_abs}
\end{figure*}

The images of the semi-major and semi-minor source axes under the linearised lens mapping given by Equation~\eqref{eq:A3} are then obtained as
\begin{align}
\boldsymbol{x}(a_\mathrm{s}) &= \mathbfss{A}^{-1} \cdot \left( \begin{matrix} \cos(2\vartheta_\mathrm{s}) & -\sin(2\vartheta_\mathrm{s}) \\ \sin(2\vartheta_\mathrm{s}) &\phantom{-} \cos(2\vartheta_\mathrm{s}) \end{matrix} \right) \cdot \left( \begin{matrix} a_\mathrm{s} \\ 0 \end{matrix} \right) \nonumber \\ 
&= \dfrac{a_\mathrm{s}}{\rm{det}(\mathbfss{A})} \left( \begin{matrix} \cos(2\vartheta_\mathrm{s}) + |\boldsymbol{g}| \cos(2\varphi - 2\vartheta_\mathrm{s}) \\ \sin(2\vartheta_\mathrm{s}) + |\boldsymbol{g}| \sin(2\varphi - 2\vartheta_\mathrm{s}) \end{matrix} \right) \;,
\label{eq:a} \\
\boldsymbol{x}(b_\mathrm{s}) &= \mathbfss{A}^{-1} \cdot \left( \begin{matrix} \cos(2\vartheta_\mathrm{s}) & -\sin(2\vartheta_\mathrm{s}) \\ \sin(2\vartheta_\mathrm{s}) & \phantom{-}\cos(2\vartheta_\mathrm{s}) \end{matrix} \right) \cdot \left( \begin{matrix}  0 \\ b_\mathrm{s} \end{matrix} \right) \nonumber \\
&= \dfrac{b_\mathrm{s}}{\rm{det}(\mathbfss{A})} \left( \begin{matrix} -\sin(2\vartheta_\mathrm{s}) + |\boldsymbol{g}| \sin(2\varphi - 2\vartheta_\mathrm{s}) \\ \phantom{-}\cos(2\vartheta_\mathrm{s}) - |\boldsymbol{g}| \cos(2\varphi - 2\vartheta_\mathrm{s}) \end{matrix} \right)
\;.
\label{eq:b}
\end{align}
Denoting the lengths of the first and the second vector in the image plane by $a$ and $b$ respectively, we determine the axis ratio of images as 
\begin{equation}
\dfrac{b}{a} = \dfrac{b_\mathrm{s}}{a_\mathrm{s}} \; \sqrt{\dfrac{(1-|\boldsymbol{g}|)^2 + 4 |\boldsymbol{g}| \sin^2 (2 \vartheta_\mathrm{s}-\varphi)}{(1-|\boldsymbol{g}|)^2 + 4 |\boldsymbol{g}| \cos^2 (2 \vartheta_\mathrm{s}-\varphi)}} \;.
\label{eq:axis_ratio}
\end{equation}
Then, the orientation angle $\vartheta$ of the image quadrupole with respect to the $x_1$-axis based on the components of the vector in Equation~\eqref{eq:a} is given by
\begin{equation}
\tan (2\vartheta) = \tan\left( \dfrac{x_2(a_\mathrm{s})}{x_1(a_\mathrm{s})} \right) = \dfrac{\sin(2\vartheta_\mathrm{s}) + |\boldsymbol{g}| \sin(2\varphi - 2\vartheta_\mathrm{s})}{\cos(2\vartheta_\mathrm{s}) + |\boldsymbol{g}| \cos(2\varphi - 2\vartheta_\mathrm{s})} \;.
\label{eq:orientation}
\end{equation}
Equations~\eqref{eq:axis_ratio} and \eqref{eq:orientation} clearly show the degeneracies between the unobservable axis ratio and orientation angle of the source and the local lens properties when only the quantities on the left-hand side can be observed. 
Assuming the intrinsic source properties to be negligible, we obtain 
\begin{align}
\dfrac{b}{a} &= \sqrt{\dfrac{(1-|\boldsymbol{g}|)^2 + 4 |\boldsymbol{g}| \sin^2 (\varphi)}{(1-|\boldsymbol{g}|)^2 + 4 |\boldsymbol{g}| \cos^2 (\varphi)}} \;, \label{eq:axis_ratio2} \\
\tan (2\vartheta) &= \dfrac{2 |\boldsymbol{g}| \sin(\varphi) \cos(\varphi)}{1 - |\boldsymbol{g}| + 2 |\boldsymbol{g}| \cos^2(\varphi)} \;. \label{eq:orientation2}
\end{align}
As only the \emph{relative} orientation between $\vartheta$ and $\varphi$ is of interest, we can transform into a coordinate system where the reduced shear is oriented in positive or negative $x_1$ direction, i.e. $\varphi = 0$ or $\varphi = \pi/2$. 
Then, Equation~\eqref{eq:axis_ratio2} shows that the semi-major and semi-minor axes are scaled in the presence of (reduced) shear, reaching the maximum and minimum axis ratios for $\varphi = \pi/2$ and $\varphi = 0$, respectively. 
Equation~\eqref{eq:orientation2} conveniently yields $\vartheta = 0$ for both cases. 
For $\varphi - \vartheta = 0$, both are aligned, for $\varphi - \vartheta = \pi/2$, the orientation angles are orthogonal, meaning that shear and image quadrupole directions are opposite to each other. 
Section~\ref{sec:application} will further detail the influence of source properties  on the reconstruction of $\boldsymbol{g}$ close to the critical curve. 

\section{Example cases}
\label{sec:application}

\subsection{Evaluation measures}
\label{sec:evaluation_measures}

Our observational analyses on galaxy-cluster-scale and galaxy-scale gravitational lenses, \cite{bib:Wagner_cluster}, \cite{bib:Griffiths}, \cite{bib:Wagner_quasar}, revealed that, on average, the confidence bounds on the local reduced shear components that we can obtain with our observation-based approach are about 10\% of the reduced shear values. 
These results imply that any change within these confidence bounds caused by intrinsic source properties or by small scale excess mass densities will remain unnoticed at the current resolution and precision of our method. 

In the following, we assume that we are given shear or reduced shear maps either of simulated gravitational lenses or of global model-based lens reconstructions. 
We use the provided maps to calculate the derivatives in $x_1$- and $x_2$-directions, abbreviated as $\partial_1$ and $\partial_2$, respectively to obtain the third order derivatives of the lensing potential. 
To do so, we employ the \texttt{MATLAB} function \texttt{gradient}\footnote{\url{https://de.mathworks.com/help/matlab/ref/gradient.html}}. 
Then, we determine
\begin{equation}
\left| \dfrac{\partial_i \gamma_j(\boldsymbol{x})}{\gamma_j(\boldsymbol{x})} \right| < 0.1 \;, \quad i,j=1,2 \;,
\label{eq:g_ratio}
\end{equation}
in which $\partial_i$ is calculated in units of $\left[ \mbox{1/px} \right]$ and $\gamma_j$ is determined for each pixel to make all units match.
Dividing numerator and denominator by $1-\kappa(\boldsymbol{x})$, the ratio is equal to the ratio of reduced shears, i.~e.~the local lens property determined in Equation~\eqref{eq:local_lens_props}. 
In addition, the ratio is constructed to be invariant over all source redshifts because the transformation from $z_\mathrm{s}=9$ to another $z_\mathrm{s}$ only amounts to a scaling. 

As some local lens properties showed tighter confidence bounds than 10\%, we also investigate 2\% as bound on the right-hand side of Equation~\eqref{eq:g_ratio}. 
Plotting the ratio on the left-hand side of Equation~\eqref{eq:g_ratio} instead of its absolute value, the sign differences between the shear and its derivatives in the regions with $\gamma_j \approx 0$ become visible as well. 

\begin{figure*}
\centering
\includegraphics[width=0.45\textwidth]{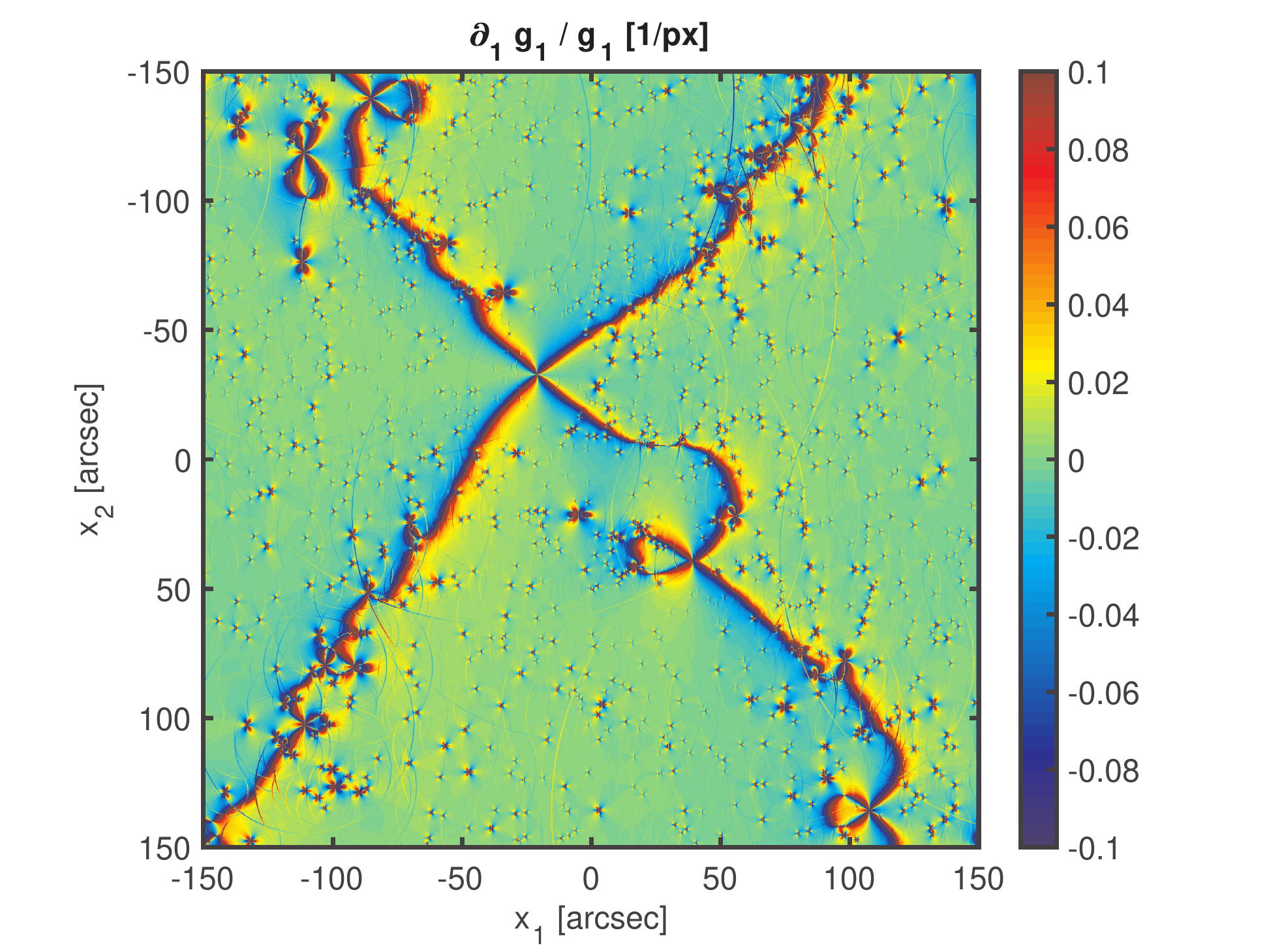} \hspace{-4ex}
\includegraphics[width=0.45\textwidth]{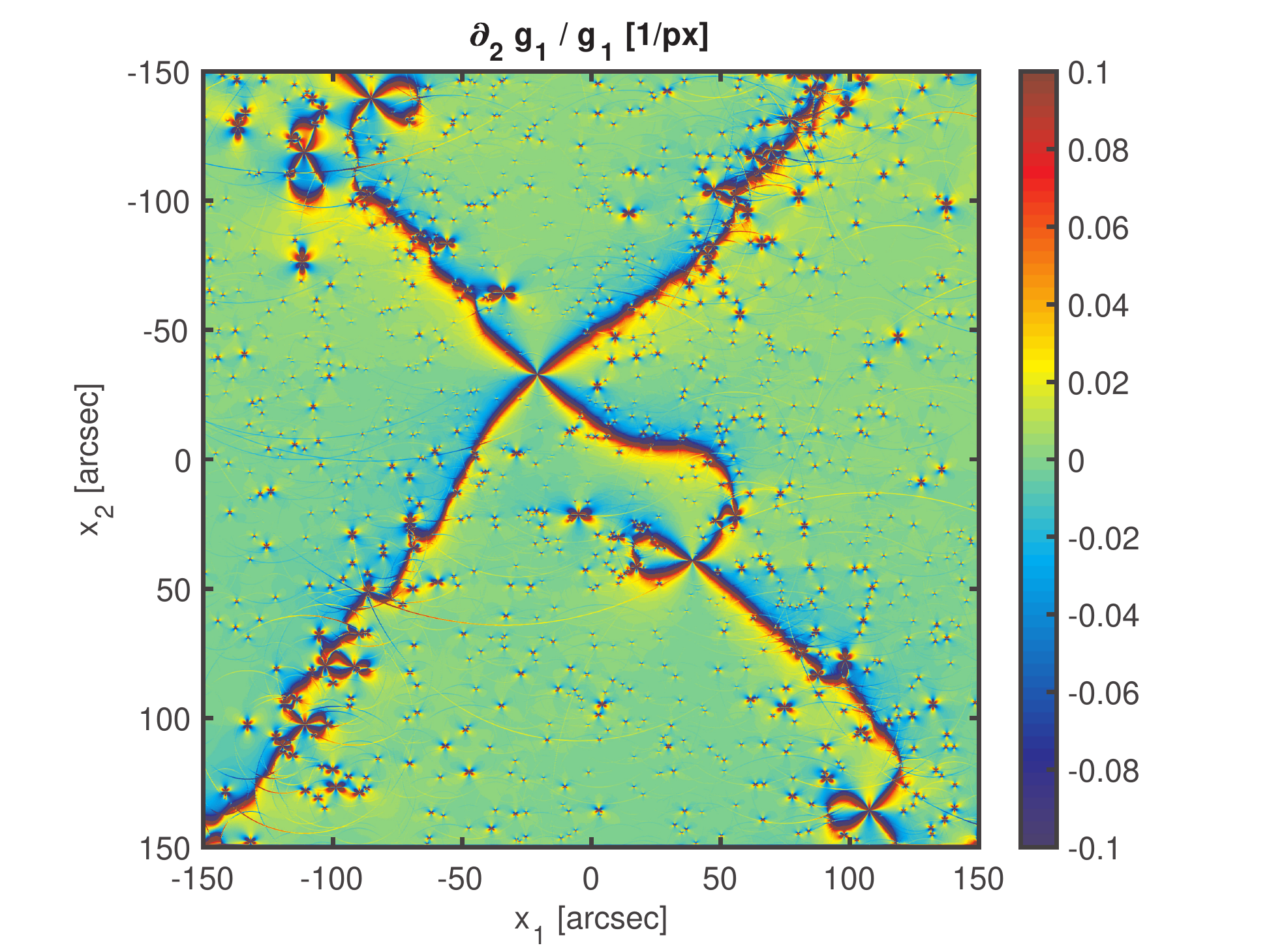}
\includegraphics[width=0.45\textwidth]{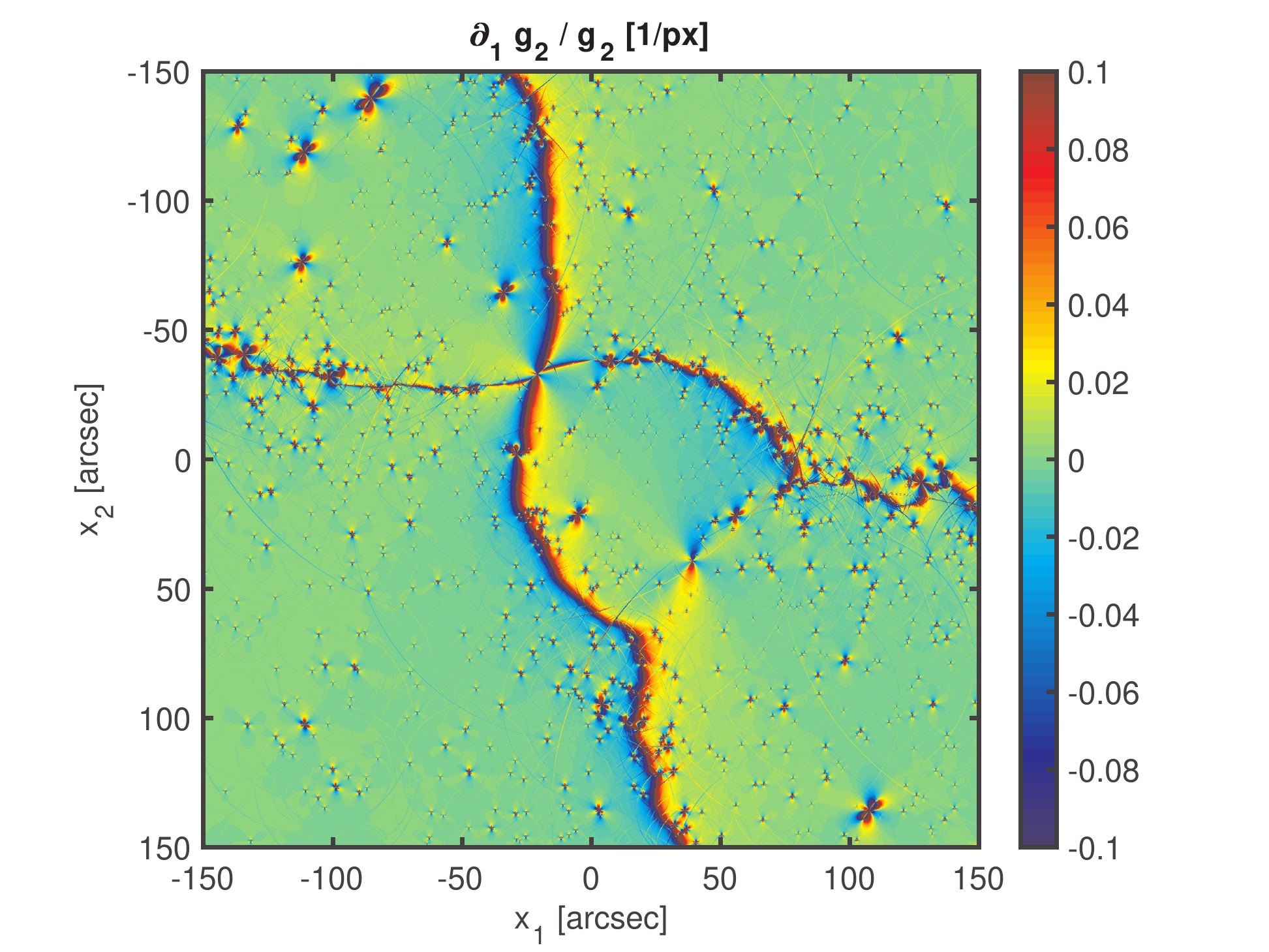} \hspace{-4ex}
\includegraphics[width=0.45\textwidth]{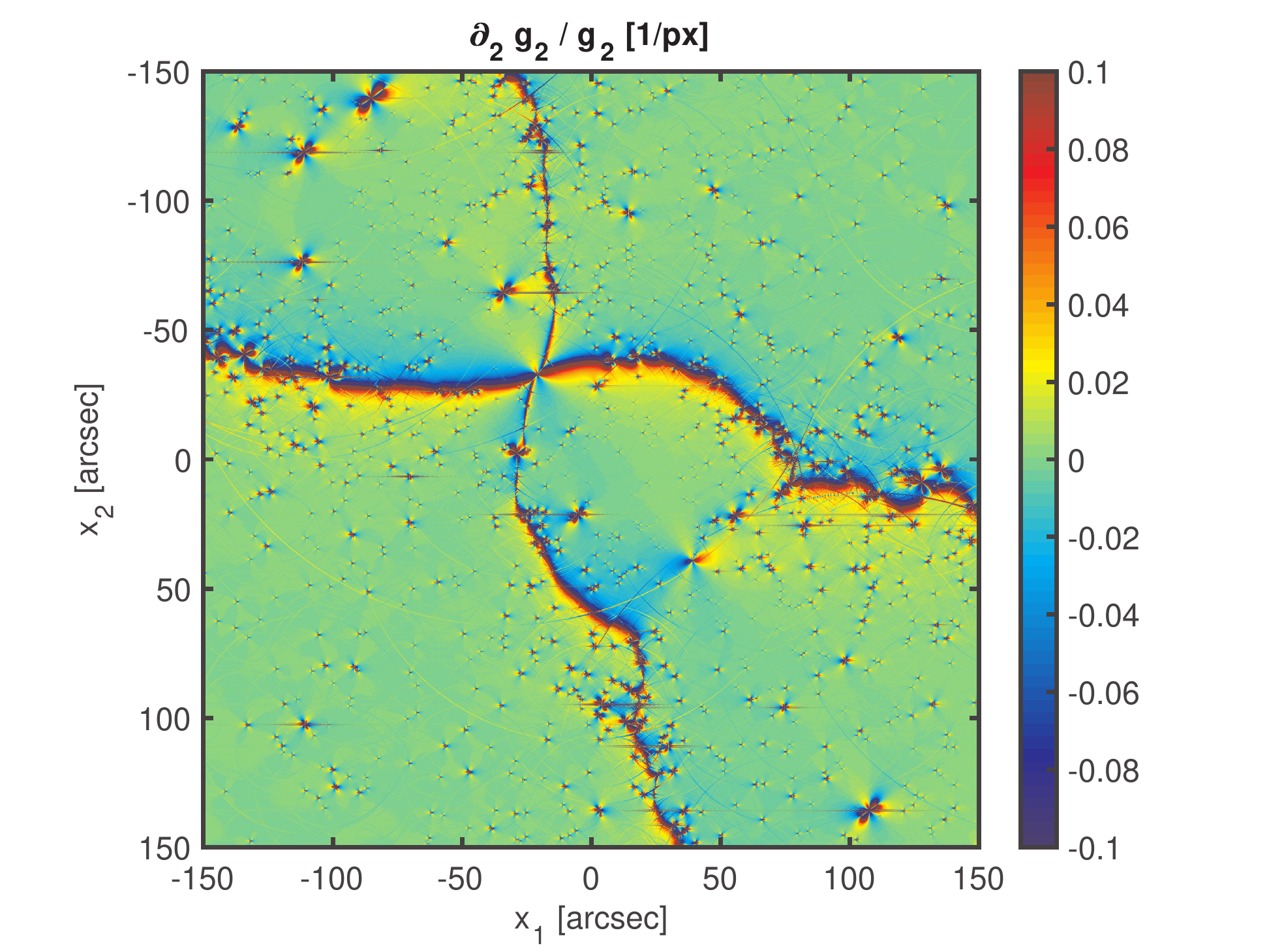}
\caption{Ratios of Equation~\eqref{eq:g_ratio} of the simulated ARES galaxy cluster. Regions with ratios larger than 10\% are plotted in red and blue. Negative values indicate regions where the shear and its derivative have opposite signs, located close to regions where the respective reduced shear component is zero. Broad green areas mark pixels with negligible higher-order lens distortions because current relative confidence bounds on the reduced shear are of the order of 10\%.}
\label{fig:ares_dg_g}
\end{figure*}

To find the region around the critical curve in which the source properties are negligible to infer local lens properties, we identify the areas in the image plane around $|\boldsymbol{g}| = 1$ with
\begin{equation}
1 - \delta_g < |\boldsymbol{g}| < 1 + \delta_g
\label{eq:cc_condition}
\end{equation}
with $\delta_g=0.1$ and $\delta_g=0.2$. 
These two values are selected because they cover the range of $|\boldsymbol{g}|$ in Fig.~\ref{fig:alignment_abs} and show how quickly $|\boldsymbol{g}|$ decreases for the different example lenses in Sections~\ref{sec:simulated_example} and \ref{sec:lenstool_example}.
As the axis ratio of the source quadrupole is directly proportional to the one in the observed image according to Equation~\eqref{eq:axis_ratio}, we only have to calculate the maximum and minimum influence of the alignment between source and (reduced) shear orientations. 
For a perfect alignment between source and shear, $\vartheta_\mathrm{s} = \varphi$, the shear further increases the length of the semi-major axis and decreases the length of the semi-minor axis of the source quadrupole. 
If the shear and the source quadrupole orientations are orthogonal $\vartheta_\mathrm{s} = \varphi - \pi/2$, the semi-minor axis is stretched and the semi-major axis decreased, such that an elliptical source will appear rounder after being lensed. 

Since most data analyses, e.g. detailed in \cite{bib:Bartelmann}, employ the complex image and source ellipticities, $\boldsymbol{\epsilon} = |\boldsymbol{\epsilon}| \exp(2i\vartheta)$ and $\boldsymbol{\epsilon}_\mathrm{s} = |\boldsymbol{\epsilon}_\mathrm{s}| \exp(2i\vartheta_\mathrm{s})$, and also a complex notation of reduced shear, $\boldsymbol{g} = |\boldsymbol{g}| \exp(2i\varphi)$, we transfer Equations~\eqref{eq:axis_ratio} and \eqref{eq:orientation} into this notation to obtain
\begin{align}
\boldsymbol{\epsilon} &= \boldsymbol{g} \dfrac{1+\tfrac{|\boldsymbol{\epsilon}_\mathrm{s}|}{|\boldsymbol{g}|}\exp(2i(\vartheta_\mathrm{s}-\varphi))}{1 + |\boldsymbol{\epsilon}_\mathrm{s}||\boldsymbol{g}|\exp(2i(\vartheta_\mathrm{s}-\varphi))} \;, \quad &\text{for} \; \;  |\boldsymbol{g}| \le 1\;, \label{eq:eps1} \\
\boldsymbol{\epsilon} &= \dfrac{1}{\boldsymbol{g}^\star} \dfrac{1+|\boldsymbol{\epsilon}_\mathrm{s}||\boldsymbol{g}|\exp(-2i(\vartheta_\mathrm{s}-\varphi))}{1 + \tfrac{|\boldsymbol{\epsilon}_\mathrm{s}|}{|\boldsymbol{g}|}\exp(-2i(\vartheta_\mathrm{s}-\varphi))} \;, \quad &\text{for} \; \;  |\boldsymbol{g}| > 1 \label{eq:eps2} \;,
\end{align}
in which $\boldsymbol{g}^\star$ denotes the complex conjugate of $\boldsymbol{g}$.
Figure~\ref{fig:alignment_abs} shows the amplitude of the image ellipticity (left) and the relative orientation with respect to the (reduced) shear orientation (right).  
The bias we obtain when inferring the reduced shear, or ratios of derivatives of the Fermat potential, from the image quadrupole close to the critical curve strongly depends on the intrinsic source properties. 
Consequently, the approximations of the local lens properties at the critical curve as derived in \cite{bib:Wagner1} depend on the proximity of the images to the critical curve. 
The most general evaluation measure that can be provided here is thus a plot of the regions around the critical curves for which Equation~\eqref{eq:cc_condition} holds. 
As the latter is not independent of the source redshift, we will show these regions for typical source redshifts. 

Furthermore, the Taylor expansion around the critical fold point predicts the position of the critical point as the centre between the two multiple images, exploiting that the Taylor expansion around this point should be symmetric to both sides in its proximity. 
Investigating the symmetry of the regions constrained by Equation~\eqref{eq:cc_condition} to both sides of $|\boldsymbol{g}|=1$ shows the limits of this approximation.

\subsection{Simulated example cases}
\label{sec:simulated_example}

\begin{figure*}
\centering
\includegraphics[width=0.45\textwidth]{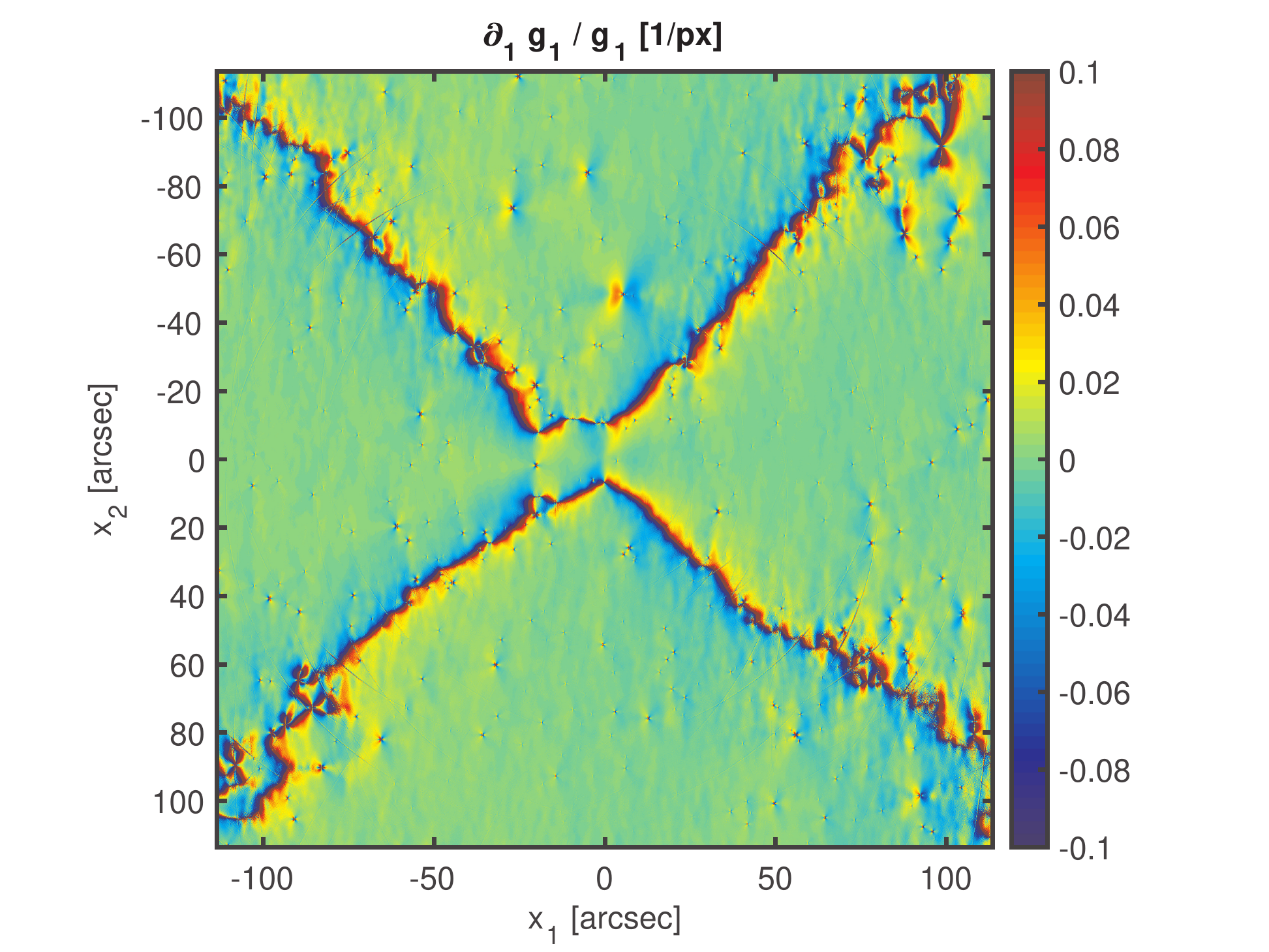} \hspace{-4ex}
\includegraphics[width=0.45\textwidth]{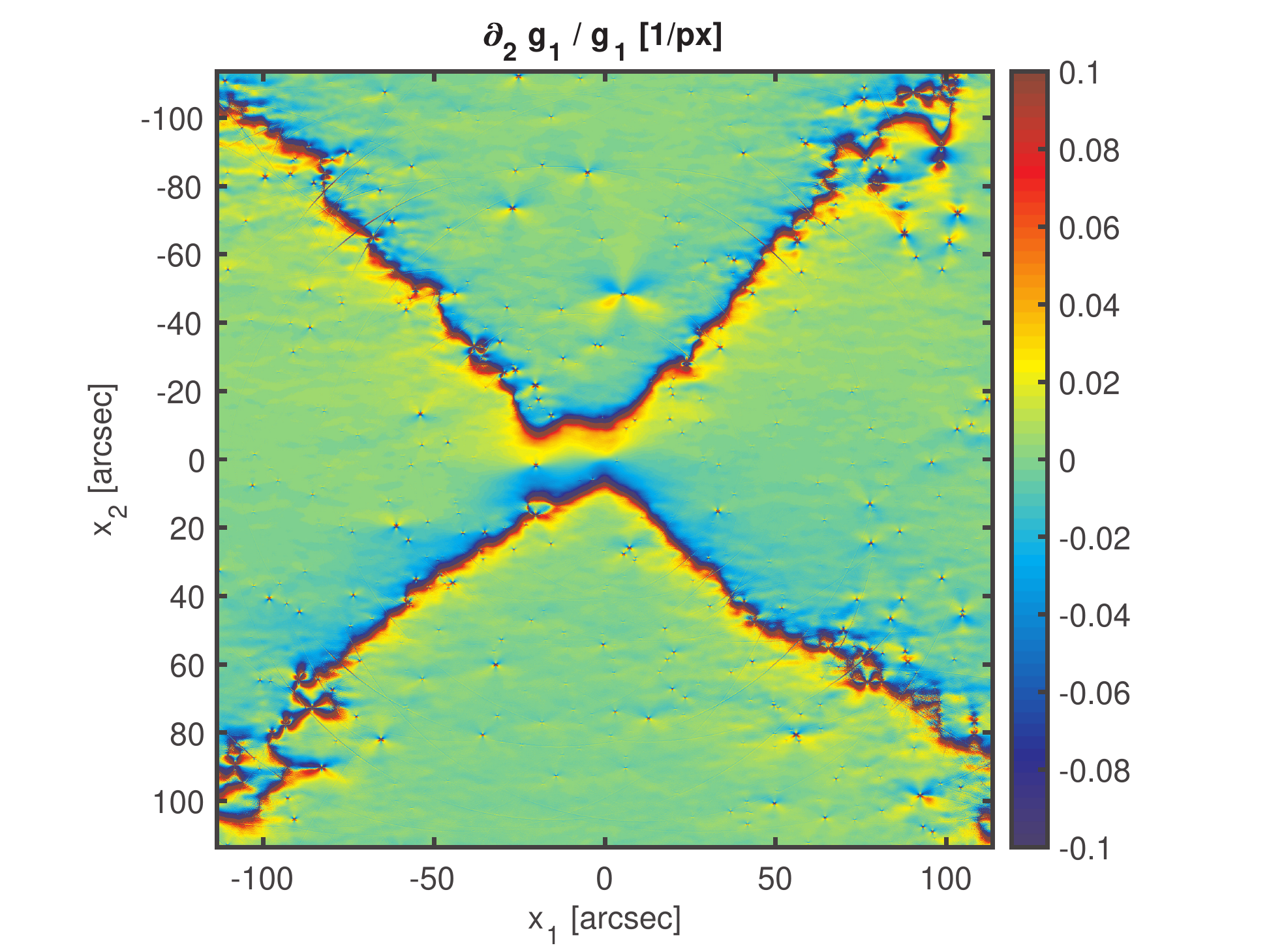}
\includegraphics[width=0.45\textwidth]{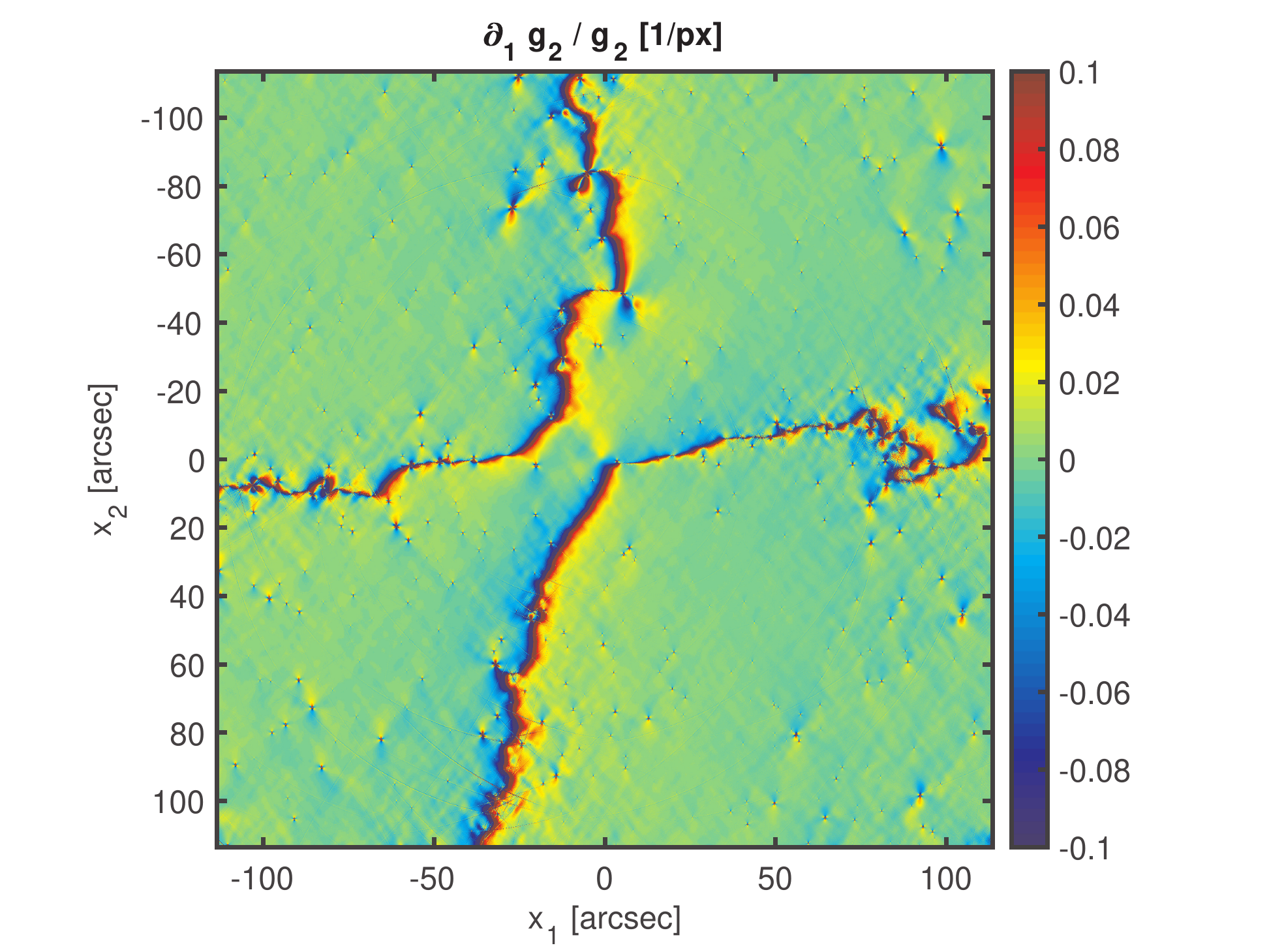} \hspace{-4ex}
\includegraphics[width=0.45\textwidth]{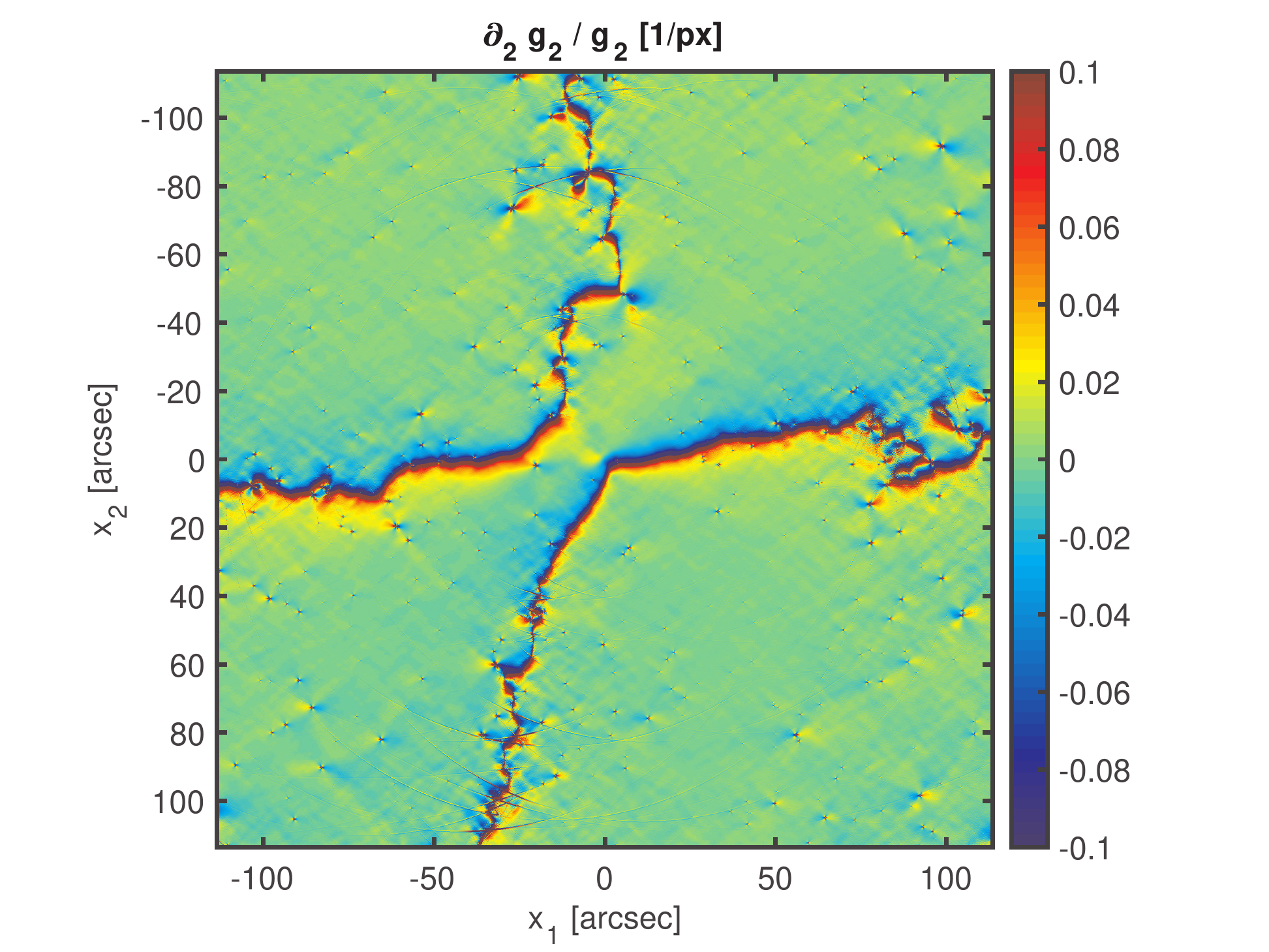}
\caption{Same as Figure~\ref{fig:ares_dg_g} for the HERA simulation.}
\label{fig:hera_dg_g}
\end{figure*}

\begin{table*}
\begin{center}
\begin{tabular}{ccccccccc}
\hline 
Case & Area & Px & Mis & Good mis & Good mis & Substr. & Good px & Good px \\
 & & & & @10\% & @2\% & mis  & @10\% & @2\% \\
\hline
\noalign{\smallskip}
ARES & $(300'')^2$ & $2048^2$ & 242 & 210 & 106 & 6 &  91.7\% & 67.8\% \\
\noalign{\smallskip}
HERA & $(227'')^2$ & $2048^2$ & 65 & 58 & 31 & 3 & 94.2\% & 72.5\% \\
\noalign{\smallskip}
CL0024 & $(600'')^2$ & $12000^2$ &  19 & 17 & 13 & 0 & 99.1\% & 96.1\% \\
\noalign{\smallskip}
CL0024 (detail) & $(200'')^2$ & $4000^2$ &  19 & 17 & 13 & 0 & 96.3\% & 85.2\% \\
\noalign{\smallskip}
\hline
\end{tabular}
\end{center}
\vspace{-2ex}
\caption{Summary of results for the examples: name of the cluster (1st column), the area covered on the sky (2nd column), total amount of pixels (px) (3rd column), number of multiple images (mis) within the total pixel area (4th column), number of multiple images with all ratios defined in Equations~\eqref{eq:g_ratio} smaller than 0.1 (5th column), number of multiple images with all ratios smaller than 0.02 (6th column), number of multiple images for which at least one ratio is larger than 0.5 being a possible hint for lensing substructure (7th column), total amount of pixels with all ratios below 0.1 (8th column), total amount of pixels with all ratios below 0.02 (9th column).}
\label{tab:sum_stats}
\end{table*}

As mentioned in Section~\ref{sec:introduction}, we use the two simulated galaxy clusters ARES and HERA, as detailed in \cite{bib:Meneghetti}. 
Their shear and convergence maps at $z_\mathrm{s}=9$ are provided at a publicly available website\footnote{\url{http://pico.oabo.inaf.it/~massimo/Public/FF/}}.
Plotting the ratios defined in Equation~\eqref{eq:g_ratio}, Figures~\ref{fig:ares_dg_g} and \ref{fig:hera_dg_g} show that there are large areas where the leading-order lensing features in the distortion matrix completely dominate the lensing effect.
As expected, in regions with reduced shear close to zero, the higher-order terms in Equation~\eqref{eq:Taylor} become relevant.
Yet, Figures~\ref{fig:ares_dg_g} and \ref{fig:hera_dg_g}  also show that the absolute values $\partial_i g_j$, $i,j=1,2$ are only a small fraction of the reduced shear values.
Consequently, as detailed in Section~\ref{sec:no_A}, multiple images lying in these regions are optimal for source reconstruction.  

Table~\ref{tab:sum_stats} gives a statistical overview of the total amount of pixels, the multiple images and the locations of negligible higher-order terms in Equation~\eqref{eq:Taylor}. 
The multiple images are represented by the pixel of the image coordinate given in the catalogue of multiple images provided by the simulation.
The ratio of Equation~\eqref{eq:g_ratio} is determined with respect to that pixel for each multiple image.
The last two columns show that the majority of pixels do not require higher-order terms to be taken into account.
More precisely, from all pixels in HERA, over 99\% of their $\partial_i g_j$, $i,j=1,2$, have an absolute value smaller than 0.005, while over 89\% of all $g_i$ have an absolute value larger than 0.01. 
For ARES, over 96\% of all pixelwise $\partial_i g_j$, $i,j=1,2$, have an absolute value smaller than 0.005 and over 96\% of the pixelwise $g_i$ have an absolute value larger than 0.01. 
Thus, higher-order lensing effects only become relevant for percent-precision local lens properties. 

\begin{figure*}
\centering
\includegraphics[width=0.48\textwidth]{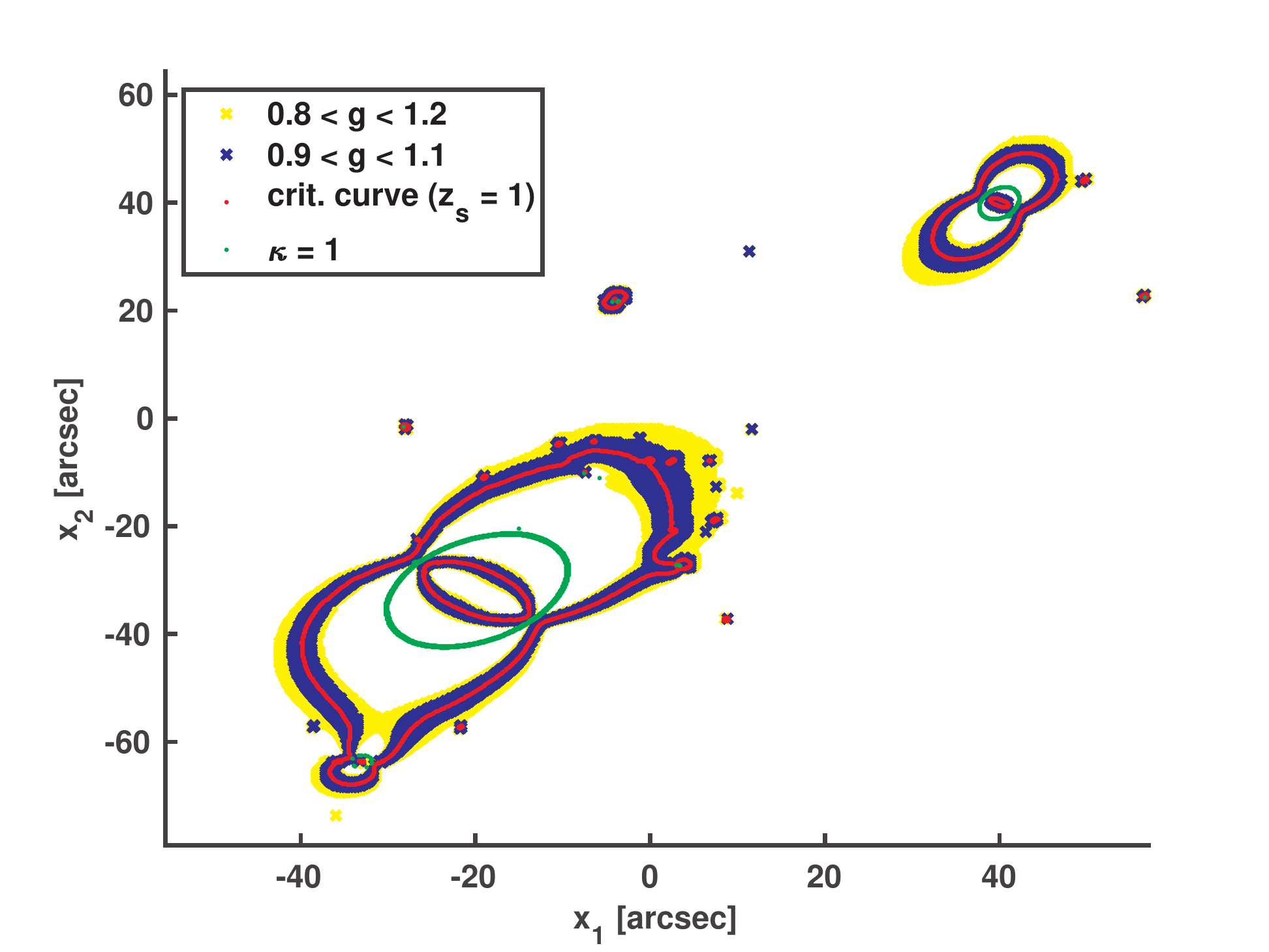} \hspace{-3ex}
\includegraphics[width=0.48\textwidth]{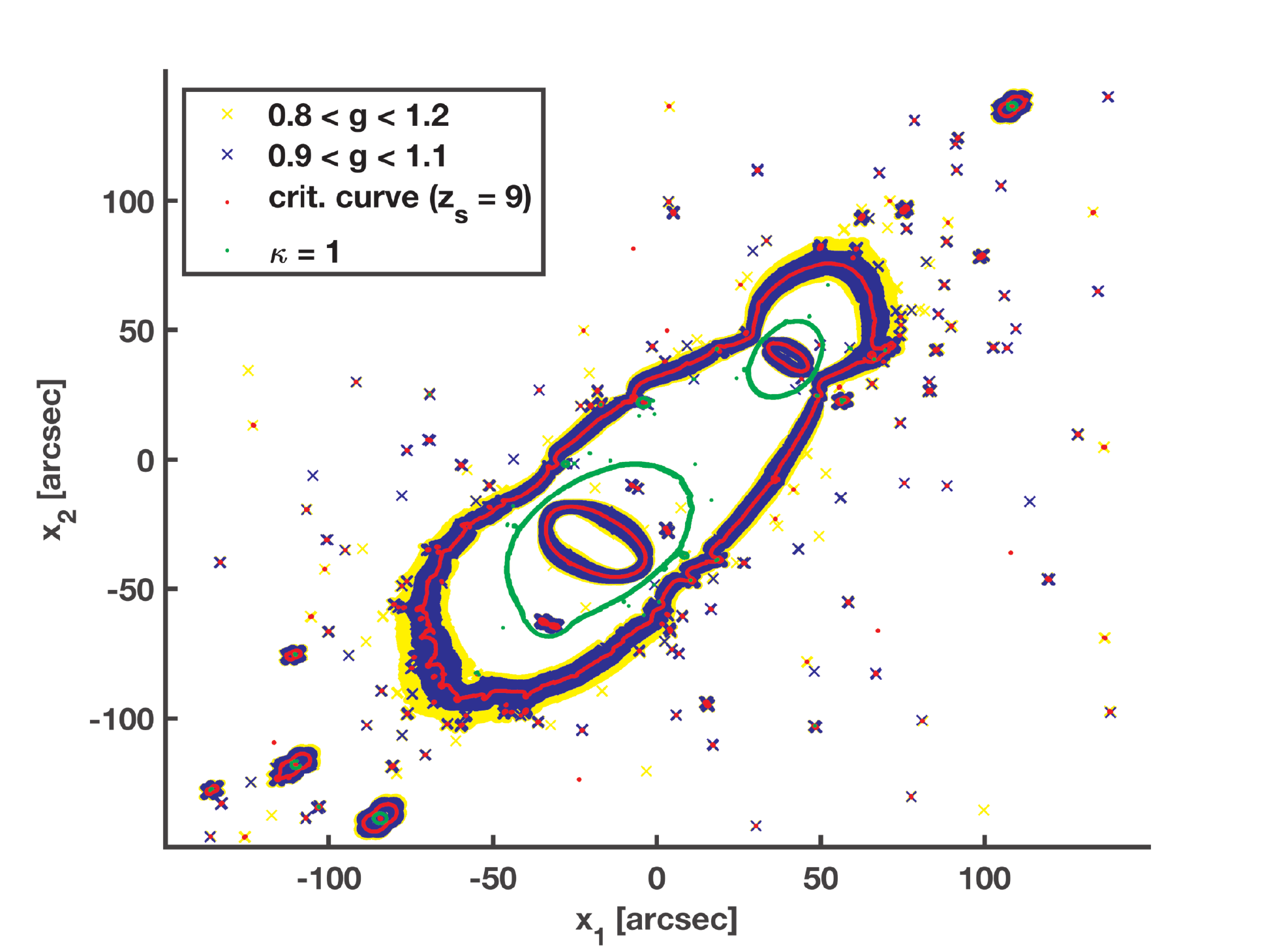}
\caption{Regions around the critical curve (red) fulfilling Equation~\eqref{eq:cc_condition} for $\delta_g = 0.1$ (blue) and $\delta_g = 0.2$ (yellow) for the ARES simulation using a source redshift of $z_\mathrm{s}=1$ (left), and $z_\mathrm{s}=9$ (right). Regions with $\kappa=1$ (green) mark areas containing multiple images that are not scaled and could therefore be used to break the mass-sheet degeneracy (see Section~\ref{sec:no_A}).}
\label{fig:ares_cc}
\end{figure*}

\begin{figure*}
\centering
\includegraphics[width=0.48\textwidth]{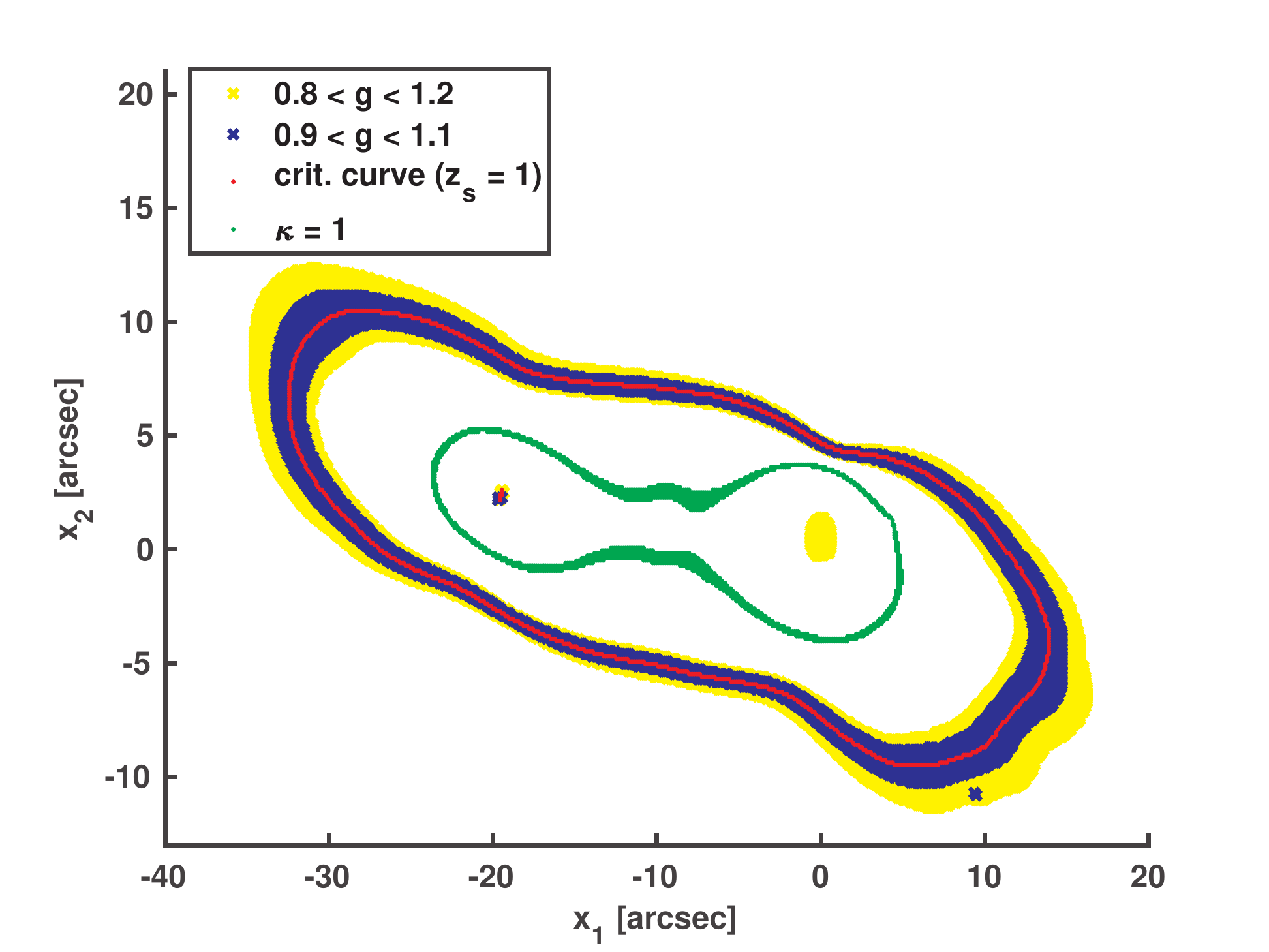} \hspace{-3ex}
\includegraphics[width=0.48\textwidth]{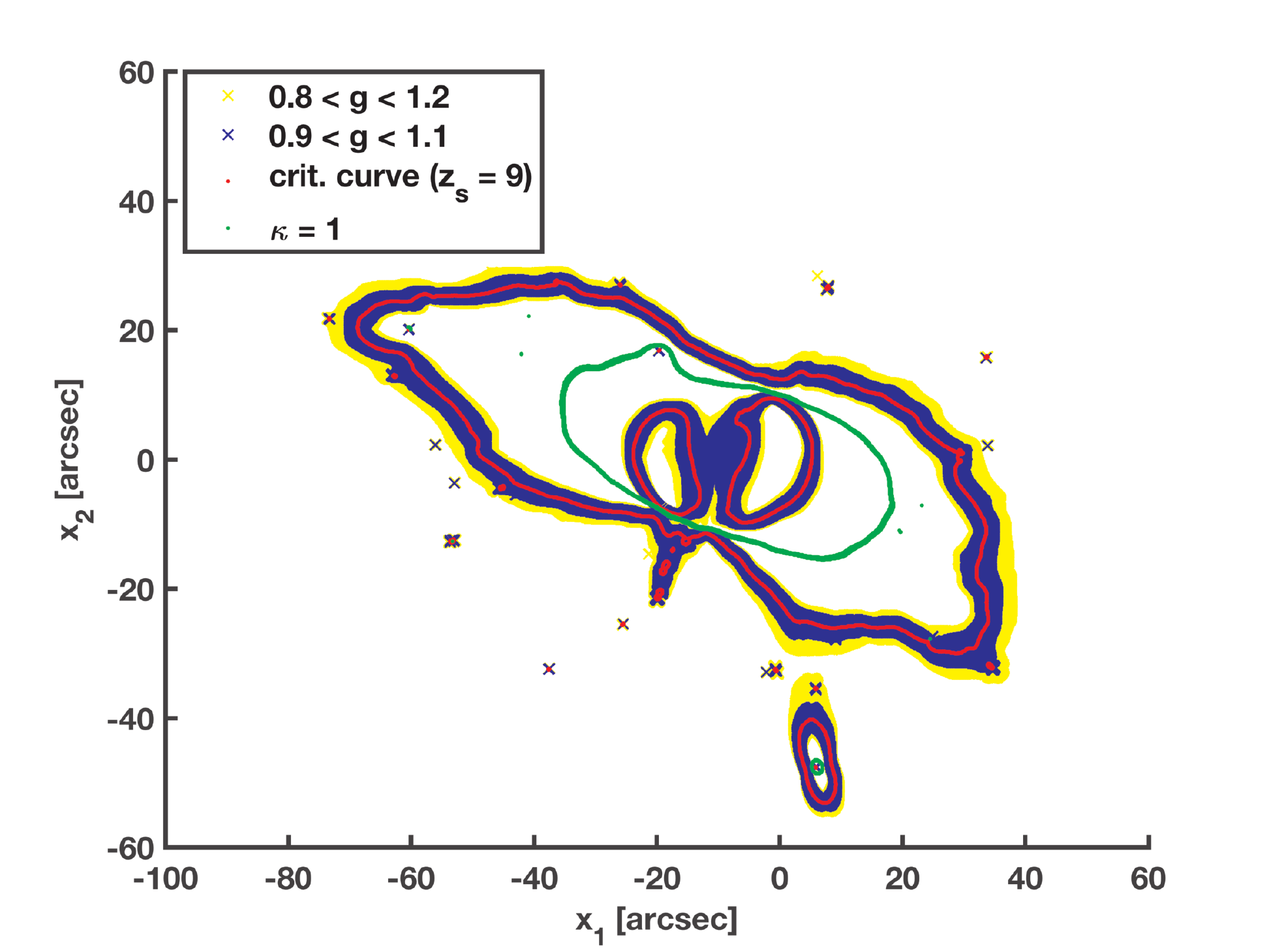}
\caption{Same as Figure~\ref{fig:ares_cc} for the HERA simulation.}
\label{fig:hera_cc}
\end{figure*}

Beyond that, we also investigated how many multiple images show at least one of the ratios in Equation~\eqref{eq:g_ratio} larger than 0.5. 
Such strong relative gradients can hint at lensing effects caused by smaller-scale structures within the lens.
However, the 6 cases for ARES and 3 cases for HERA are located in regions with very small reduced shear components, i.~e.~at places where we already expect smaller-scale lensing effects to dominate. 

We can thus conclude that, at the current level of precision in the local lens properties of Equation~\eqref{eq:local_lens_props}, the lensing effects contained in the distortion matrix clearly dominate over higher-order ones, if these simulations indeed resemble observable lensing configurations.

Plotting the regions around the critical curve fulfilling the constraints set by Equation~\eqref{eq:cc_condition} depends on the redshift of the source, so that we show the plots for $z_\mathrm{s}=1$ for typical source redshifts and $z_\mathrm{s}=9$ for high-redshift sources which are still rare but may become more abundant with data from the James Webb Space Telescope. 
Figure~\ref{fig:ares_cc} shows the plots for ARES, Figure~\ref{fig:hera_cc} the same for the HERA simulation. 

For $z_\mathrm{s}=9$, the $\delta_g=0.1$ regions of Equation~\eqref{eq:cc_condition} in the ARES simulation form a band of about 5'' width and even broader close to the cusp regions at the bottom left and top right of the plot. 
For HERA, this band is a bit smaller, but its width is also in the range of 2-4''. 
For $z_\mathrm{s}=1$, the $\delta_g=0.1$ regions for ARES are smaller and span a band of about 2''. 
Similarly, the same plot for HERA shows a reduced area of about 1'' width. 
The $\delta_g=0.1$ bands are more symmetric than the $\delta_g=0.2$ bands in both simulated galaxy clusters and the symmetry assumptions required for the approach of \cite{bib:Wagner1} are certainly met in those bands.

\begin{figure*}
\centering
\includegraphics[width=0.45\textwidth]{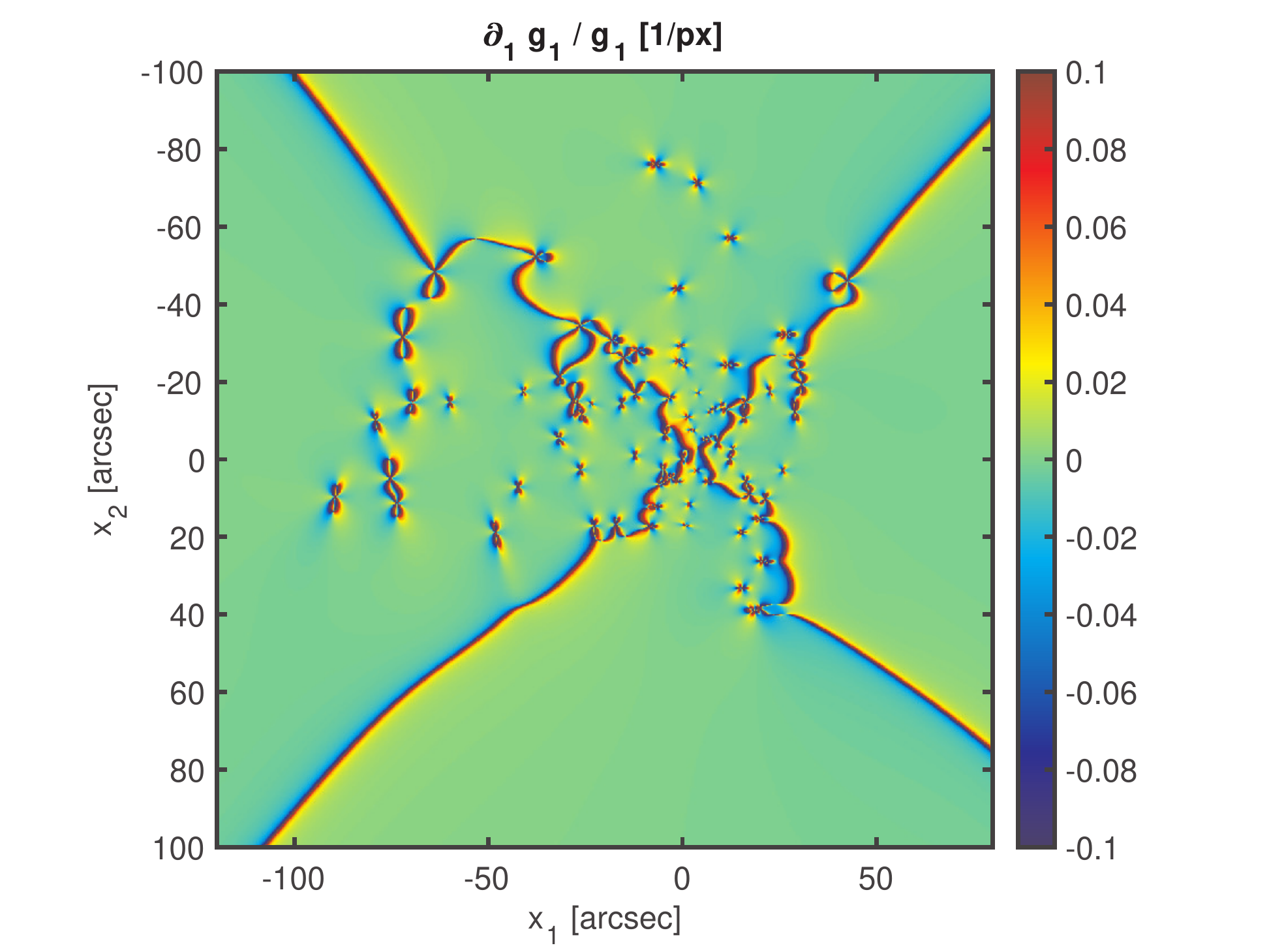} \hspace{-4ex}
\includegraphics[width=0.45\textwidth]{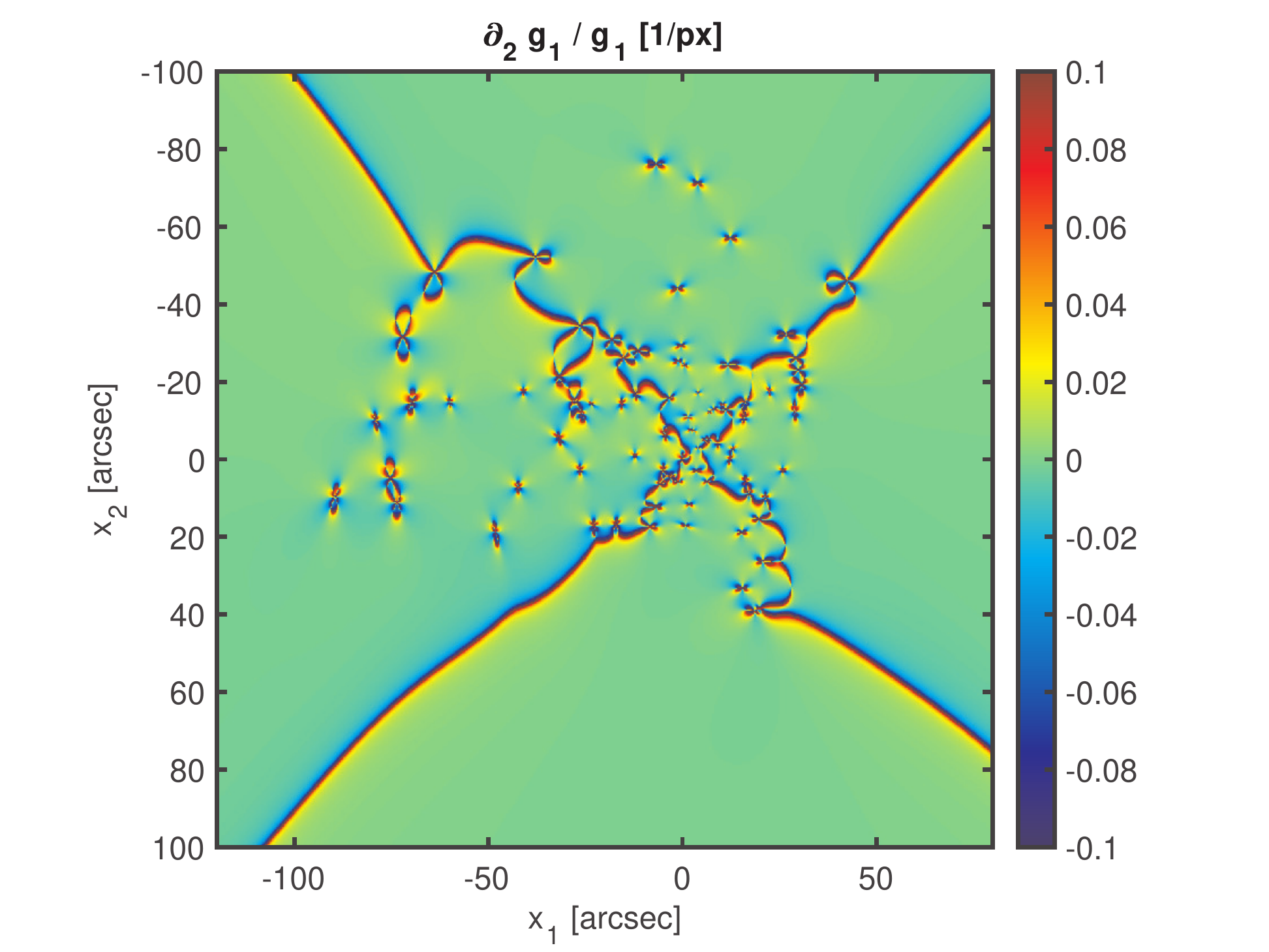}
\\[2ex]
\includegraphics[width=0.45\textwidth]{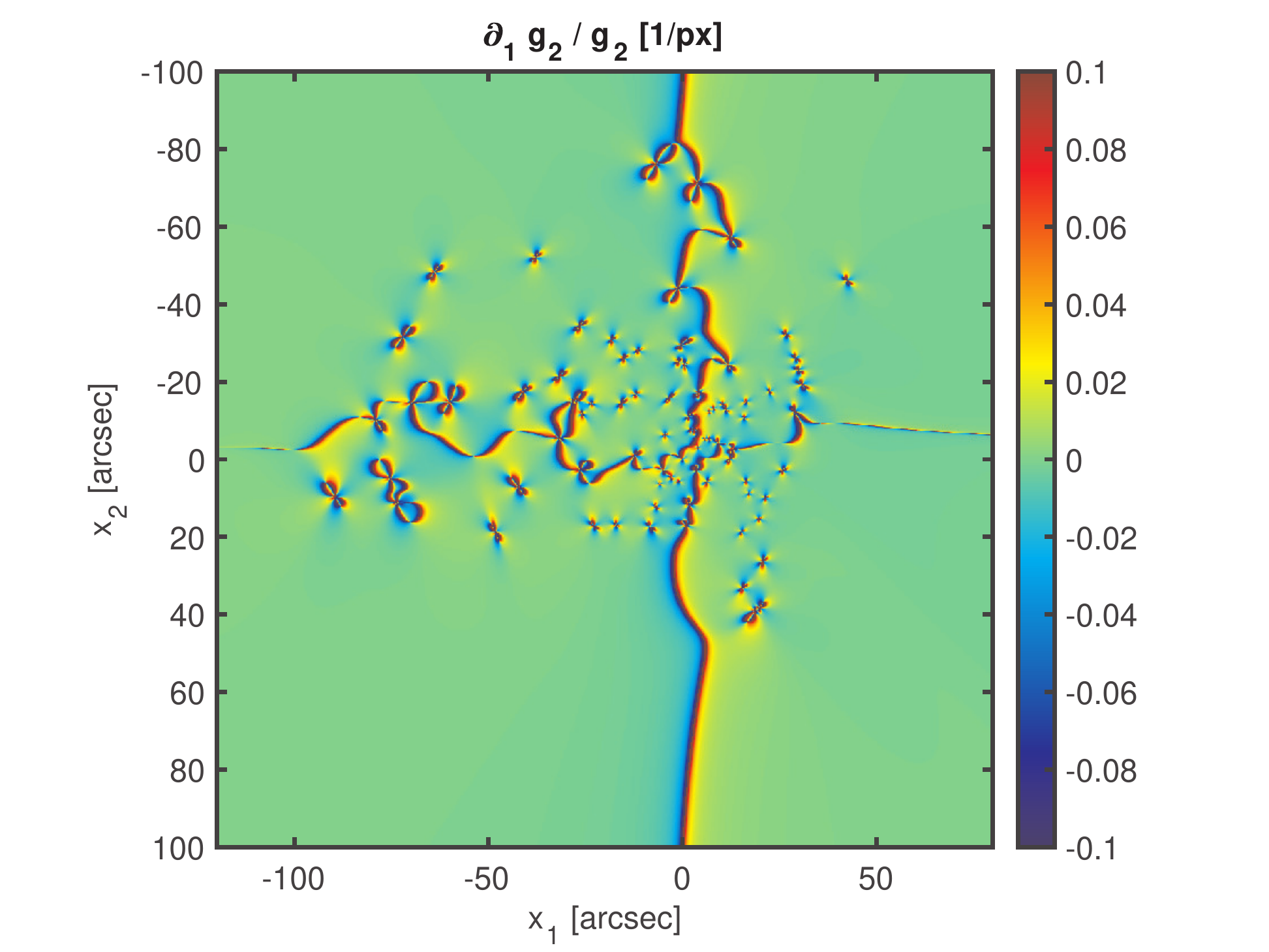} \hspace{-4ex}
\includegraphics[width=0.45\textwidth]{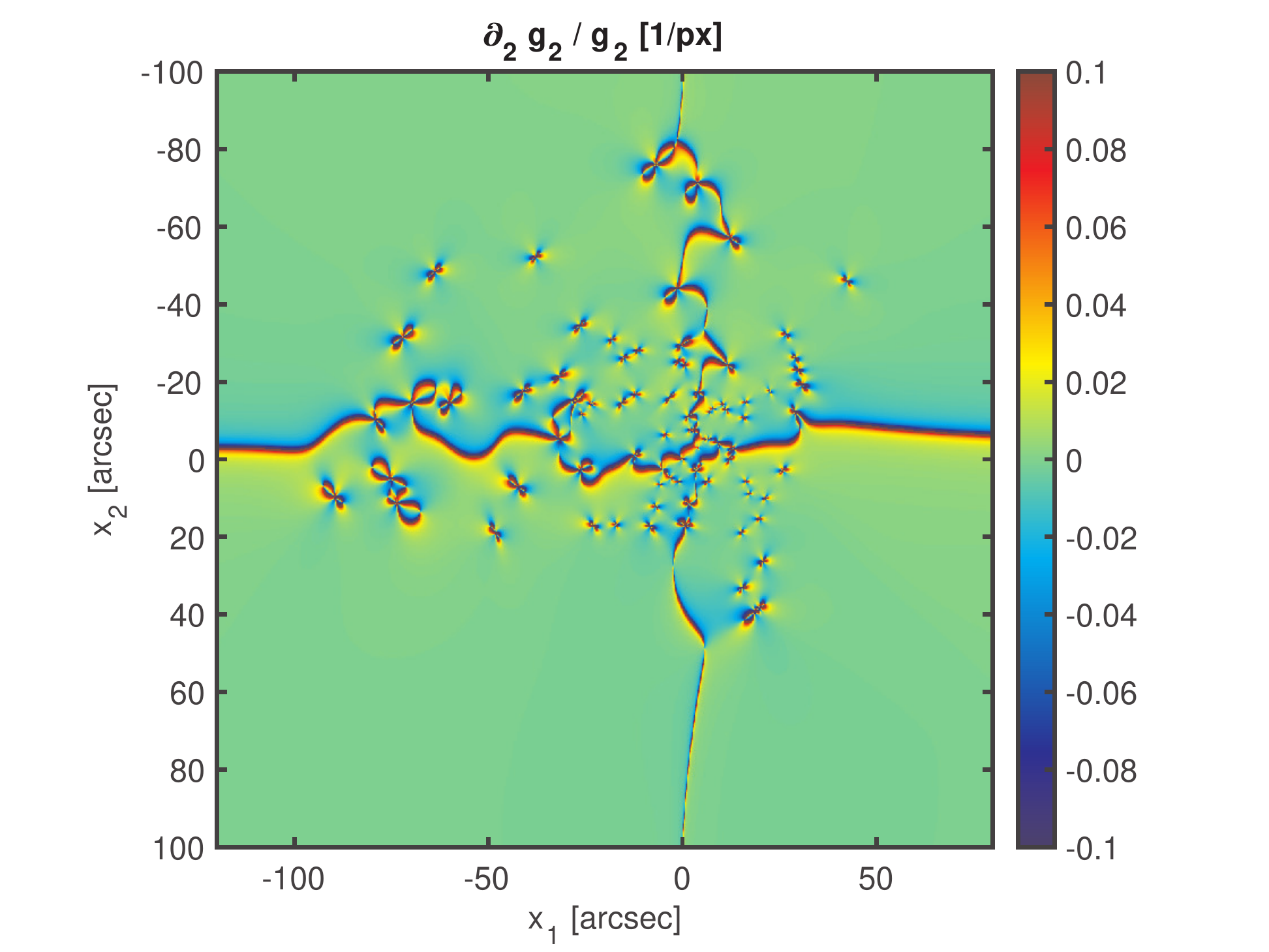}
\caption{Same as Figure~\ref{fig:ares_dg_g} for the CL0024 reduced shear maps. The plots here show the relevant detail of the total reconstructed area listed in Table~\ref{tab:sum_stats}.}
\label{fig:cl0024_dg_g}
\end{figure*}

\subsection{Lenstool lens reconstruction example case}
\label{sec:lenstool_example}

To complement the artificial lensing configuration study, we also apply the evaluation detailed in Section~\ref{sec:evaluation_measures} to the lens reconstruction as obtained by \texttt{Lenstool} for the galaxy cluster lens CL0024 in \cite{bib:Wagner_cluster}. 
Since \texttt{Lenstool} uses the light-traces-mass assumption to reconstruct the deflecting mass density profile around observed member galaxies, the reconstruction includes the 85 brightest cluster member galaxies as deflecting substructures on top of two cluster-scale pseudo-isothermal elliptical mass density profiles which model the dark matter contents. 

As Table~\ref{tab:sum_stats} shows, the reconstruction covers a larger area on the sky and also has a higher resolution than the simulations. 
Investigating the dependence of the evaluation on the area, we analyse the relevant detail around the cluster centre which is comparable to the simulated cases.
The results are summarised in the second row for this case. 
As expected, the lens reconstruction contains fewer small-scale substructures, so that the amount of pixels with negligible gradients is higher compared to the simulated clusters. 
Figure~\ref{fig:cl0024_dg_g} shows that this result is not merely achieved by the increased resolution because a comparison between the plots and the ones in Figures~\ref{fig:ares_dg_g} and \ref{fig:hera_dg_g} shows that the lens mass reconstruction is indeed smoother than the simulations. 
As detailed in \cite{bib:Wagner_cluster}, the multiple image systems used to arrive at these convergence and shear maps are well fitted by the two dark matter halos and the 85 cluster member galaxies, so that no further substructure causing large gradients is necessary to describe the observables. 
There are two (or six) multiple images that do not fulfill the thresholds of Equation~\eqref{eq:g_ratio}.
Yet, as in the simulations, these images lie again in areas with $g_i \approx 0$, $i=1,2$. 

\begin{figure}
\includegraphics[width=0.5\textwidth]{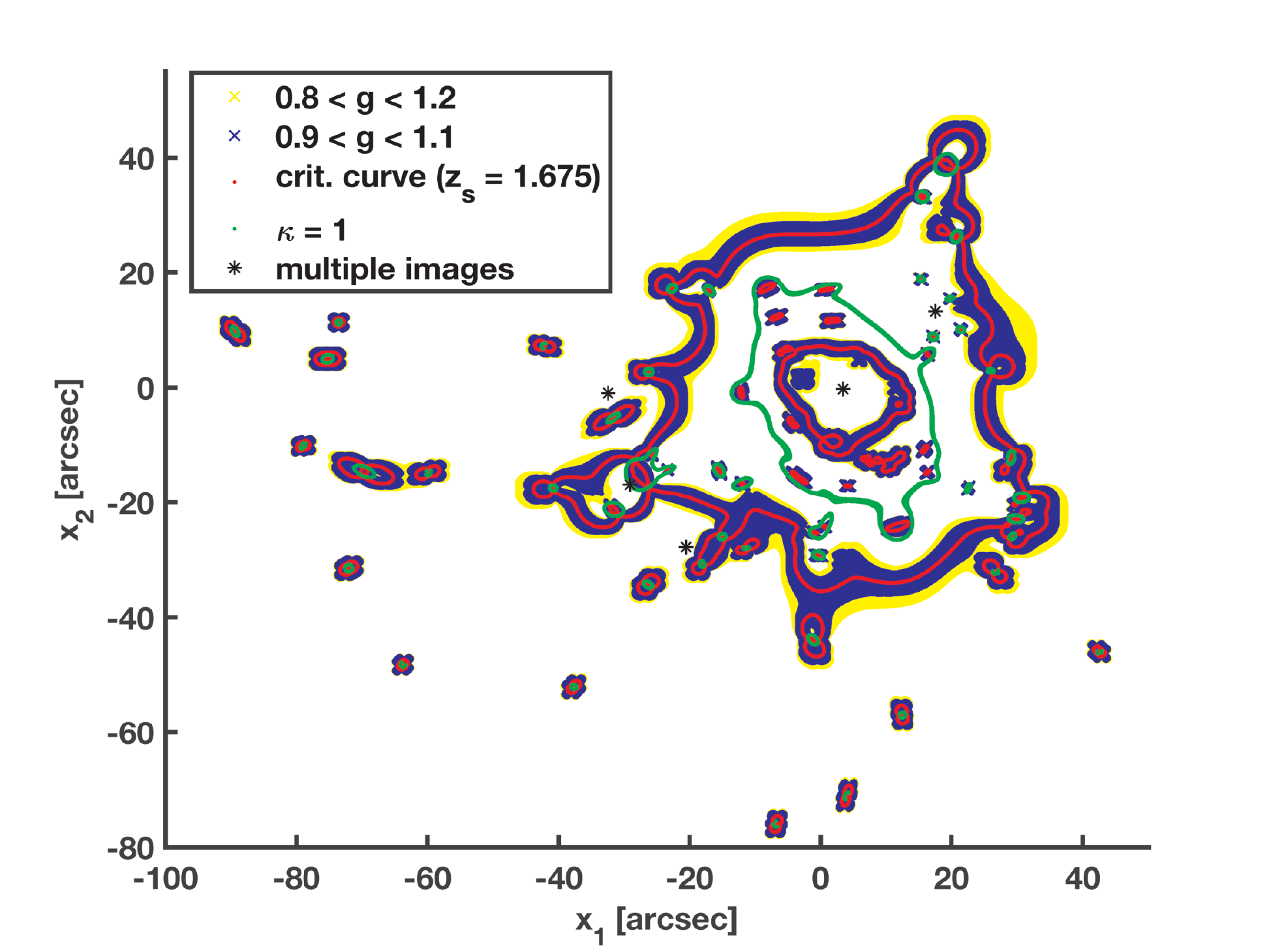}
\caption{Same as Figure~\ref{fig:ares_cc} for the convergence and shear maps reconstructed for CL0024 by \texttt{Lenstool}.}
\label{fig:cl0024_cc}
\end{figure}

Plotting the regions around the critical curve fulfilling the condition set in Equation~\eqref{eq:cc_condition} for the source redshift of the 5-image configuration at $z_\mathrm{s}=1.675$ discussed in detail in \cite{bib:Wagner_cluster}, we observe a similar result as for the simulated clusters, see Figure~\ref{fig:cl0024_cc}.
The width of $\delta_g=0.1$ regions around this critical curve in CL0024 is of the order of 2'', mostly larger, and these regions are also very symmetric around fold critical points. 
Regions with $\delta_g=0.2$ are less symmetric on both sides of the critical curves and only marginally broader than the $\delta_g=0.1$ regions for the inner critical curve, similar to the simulated cases as well. 

Figure~\ref{fig:cl0024_cc} also shows the position of the five multiple images of the $z_\mathrm{s}=1.675$ galaxy within the landscape of critical curves and lens properties. 
Based on this \texttt{Lenstool} reconstruction, the decision not to employ the approach of \cite{bib:Wagner1} to get an approximation of the critical curve from the three-image cusp configuration in the lower left of the plot was a good choice in \cite{bib:Wagner_cluster}, as Figure~\ref{fig:cl0024_cc} reveals that these images do not lie within the regions around the critical curve for which the required approximations may be valid. 

In addition, Figure~\ref{fig:cl0024_cc} reveals that image~2 (central image in the cusp configuration) lies in a region with $\kappa \approx 1$, which is also supported in a comparison to the reconstruction by \texttt{Grale} (see Figure~9 of \cite{bib:Wagner_cluster}). 
Thus, multiple images located in regions with $\kappa \approx 1$ may not be rare, given that our first analysis already contained one example. 
Yet, this image is contaminated by a lot of stray light from surrounding foreground objects such that it may be difficult to employ to break the mass-sheet degeneracy. 

\section{Conclusion}
\label{sec:conclusion}

In this part of the paper series we addressed the general problem of gravitational lensing that it is not possible to jointly reconstruct the gravitational lens and the brightness profile of an extended background source object only using the observable brightness profiles of the multiple images and the standard single-lens-plane gravitational lensing formalism. 
This inverse problem is under-constrained, such that additional assumptions about properties of the lens or the source are required to be inserted into the lensing formalism.
Employing models for the deflecting mass density or the brightness profile of the source (or both) that match the observational evidence, we obtain self-consistent solutions for the deflecting mass density profile and the brightness profile of the background source. 

Since there is no unique self-consistent solution to describe an observed multiple image configuration, we investigated which local lens properties and source properties can be inferred purely based on the lensing formalism and the observables.
We found that only \emph{relative} local lens properties between the multiple images can be constrained by mapping pairs of multiple images onto each other, as developed in \cite{bib:Wagner2}, see Equation~\eqref{eq:T_map}. 
The degeneracy between small-scale local lens properties and intrinsic source properties, Equation~\eqref{eq:deg3}, is broken for infinitesimal beams of constant surface brightness or for extended sources with known intrinsic properties. 
Furthermore, images lying in regions with $\kappa \approx 1$ can help to break the mass-sheet degeneracy, if the $\kappa \approx 1$ regions can be reliably identified. 
Vice versa, images in regions with vanishing shear are optimal starting points to reconstruct the source morphology, as they are only enlarged by the lensing effect. 

Based on the evaluation of the simulated HERA and ARES galaxy clusters and a global reconstruction of CL0024 by \texttt{Lenstool}, we also find that higher-order lensing effects like flexion are negligible for most part of the lensing regions and only become relevant in regions with vanishing shear. 
Hence, going to percent-precision local lens properties, these effects may need to be included into our approach, but are not relevant at the current precision level. 
Table~\ref{tab:sum_stats} summarises the results of this analysis. 
As we only considered pixelwise differences in convergence and reduced shear, these considerations are limited to small-scale substructures causing higher-order deflections. Yet, the method proposed here is easy to extend for any coarse-grained lens reconstruction on a larger scale, e.g.~to account for galaxy-scale lenses within the cluster lens. Yet, as mentioned at the beginning, any giant arc on galaxy-cluster or galaxy scale certainly requires higher-order deflections to be taken into account and was excluded from this analysis a priori.

Our approach to determine local lens properties at the critical curve also suffers from the degeneracy between lens and intrinsic source properties, to leading order from a source ellipticity deviating from 1. 
Yet, the influence of the source decreases quickly with decreasing distance to the critical curve. 
Figure~\ref{fig:alignment_abs} visualises the impact of a source ellipticity on the inference of reduced shear from a measured image ellipticity, as determined in Equations~\eqref{eq:eps1} and \eqref{eq:eps2}. 
For reduced shear amplitudes between 0.9 and 1.1, small intrinsic ellipticities, |$\boldsymbol{\epsilon}_\mathrm{s}| \le 0.2$, cause deviations of less than 10\% in the inferred amplitude of the reduced shear and less than $3^\circ$ in the orientation between the observed image ellipticity and the reduced shear. 
Identifying regions with these properties close to the critical curves in the simulated galaxy clusters and in the CL0024 reconstruction, we find that they span bands of several arcseconds width.
Hence, our approximations to local lens properties at the critical curve obtained from multiple images within these distances to the critical curves actually come close to the true value, see e.g.~\cite{bib:Griffiths} for an example.  

The results in this part of the paper series finally conclude the study of all possible single-lens-plane degeneracies and we will now continue on improving the ways to break them. 
For instance, the next part of the series will investigate the change of the local lens properties over a range of observed wavelengths, which can contribute to further constrain the emission profile of the source or the dust content of the lens, as lensing is an achromatic effect. 

%

\begin{acknowledgements}
I would like to thank Jori Liesenborgs, Massimo Meneghetti, Julian Merten and Liliya L.~R.~Williams for helpful discussions and comments.
\end{acknowledgements}

\bibliographystyle{aa}
\bibliography{aa}

\appendix

\section{Derivation of Equation~(14)}
\label{app:T} 

To arrive at Equation~\eqref{eq:T_map}, we start from Equation~\eqref{eq:T} and insert the distortion matrix as defined in Equation~\eqref{eq:A2} for each image $I$ and $J$.
The fully expanded expressions of each entry of the matrix $\mathbfss{T}_{IJ}$ can be found in \cite{bib:Wagner2}, Equations~(8)-(11):
\begin{align}
\mathbfss{T}^{(I,J)}_{11} &=  \dfrac{1-\kappa_I}{1-\kappa_J}  \dfrac{\left(1-g_{I,1}\right)\left( 1 + g_{J,1} \right)-g_{I,2} g_{J,2}}{1-\left(g_{J,1}\right)^2 - \left(g_{J,2}\right)^2} \;, \label{eq:T11} \\
\mathbfss{T}^{(I,J)}_{12} &= \dfrac{1-\kappa_I}{1-\kappa_J} \dfrac{\left(1+g_{I,1}\right) g_{J,2} - \left( 1 + g_{J,1} \right)g_{I,2}}{1-\left(g_{J,1}\right)^2 - \left(g_{J,2}\right)^2}  \;, \label{eq:T12}\\
\mathbfss{T}^{(I,J)}_{21} &= \dfrac{1-\kappa_I}{1-\kappa_J} \dfrac{\left(1-g_{I,1}\right) g_{J,2} - \left( 1 - g_{J,1} \right)g_{I,2}}{1-\left(g_{J,1}\right)^2 - \left(g_{J,2}\right)^2}  \;, \label{eq:T21} \\
\mathbfss{T}^{(I,J)}_{22} &= \dfrac{1-\kappa_I}{1-\kappa_J} \dfrac{\left(1+g_{I,1}\right)\left( 1 - g_{J,1} \right)-g_{I,2} g_{J,2}}{1-\left(g_{J,1}\right)^2 - \left(g_{J,2}\right)^2}  \;. \label{eq:T22}
\end{align}
For the sake of clarity, the subscripts on the transformation now denote the matrix entries of $\mathbfss{T}_{IJ}$ and the images are indicated by the $(I,J)$ superscript. 
Then, instead of solving for individual $f$ and $\boldsymbol{g}$ as in Equation~\eqref{eq:local_lens_props}, we rewrite the transformation in terms of $\kappa_I - \kappa_J$ and $\boldsymbol{g}_I - \boldsymbol{g}_J$ to arrive at Equation~\eqref{eq:T_map}. 

\end{document}